\documentclass{jfm}

\usepackage{multirow}
\usepackage{graphicx}
\usepackage{natbib}
\usepackage{epstopdf}
\usepackage{amsmath}
\usepackage{flushend}
\usepackage{rotating}

\usepackage{ amssymb, amsfonts}

\usepackage{xcolor}

\ifCUPmtlplainloaded \else
  \checkfont{eurm10}
  \iffontfound
    \IfFileExists{upmath.sty}
      {\typeout{^^JFound AMS Euler Roman fonts on the system,
                   using the 'upmath' package.^^J}%
       \usepackage{upmath}}
      {\typeout{^^JFound AMS Euler Roman fonts on the system, but you
                   dont seem to have the}%
       \typeout{'upmath' package installed. JFM.cls can take advantage
                 of these fonts,^^Jif you use 'upmath' package.^^J}%
      }
  \else
  \fi
\fi

\ifCUPmtlplainloaded \else
  \checkfont{msam10}
  \iffontfound
    \IfFileExists{amssymb.sty}
      {\typeout{^^JFound AMS Symbol fonts on the system, using the
                'amssymb' package.^^J}%
       \usepackage{amssymb}%

      }{}
  \fi
\fi

\ifCUPmtlplainloaded \else
  \IfFileExists{amsbsy.sty}
    {\typeout{^^JFound the 'amsbsy' package on the system, using it.^^J}%
     \usepackage{amsbsy}}
    {}
\fi

\newsavebox{\astrutbox}
\sbox{\astrutbox}{\rule[-5pt]{0pt}{20pt}}

\newcommand{\dd}{\ensuremath{\mathrm{d}}}
\newcommand{\id}{\ensuremath{\hphantom{.}\mathrm{d}}}
\newcommand{\f}[2]{\ensuremath{\frac{#1}{#2}}} 

 \mathchardef\mhyphen="2D
\newcommand{\pf}[2]{\ensuremath{\frac{\partial{#1}}{\partial{#2}}}}

\newcommand{\solidLine}[1]{\protect\raisebox{0.8ex}{\color{#1}\linethickness{0.5mm}\line(1,0){0.6}}}
\newcommand{\dashLine}[1]{\protect\raisebox{0.8ex}{\color{#1}\linethickness{0.5mm}\line(1,0){0.3}\hspace{0.15cm}\line(1,0){0.3}}}
\newcommand{\dashDotLine}[1]{\protect\raisebox{0.8ex}{\color{#1}\linethickness{0.5mm}\line(1,0){0.3}\hspace{0.15cm}\line(1,0){0.1}\hspace{0.1cm}\linethickness{0.5mm}\line(1,0){0.3}}}
\newcommand{\dotLine}[1]{\protect\raisebox{0.8ex}{\color{#1}\linethickness{0.5mm}\line(1,0){0.1}\hspace{0.1cm}\line(1,0){0.1}\hspace{0.1cm}\line(1,0){0.1}}}

\definecolor{myred2}{RGB}{179, 13, 13}
\definecolor{myred}{RGB}{255, 0, 0}
\definecolor{myblue2}{RGB}{0, 0, 153}
\definecolor{myblue}{RGB}{0, 0, 255}
\definecolor{mygreen}{RGB}{0, 102, 0}
\definecolor{myorag}{RGB}{246, 150, 38}
\definecolor{mypurp}{RGB}{73,0,146}
\definecolor{myyel}{RGB}{235,235,10}
\definecolor{LGrey}{rgb}{.5,.5,.5}

\title[Roughness effects in turbulent forced convection]{
Roughness effects in turbulent forced convection}

\author[M. MacDonald, N. Hutchins and D. Chung]
{M. MacDonald\thanks{Email address for correspondence: michael.macdonald@unimelb.edu.au},\ns
N. Hutchins
 and D. Chung
}

\affiliation{
Department of Mechanical Engineering, University of Melbourne, Victoria 3010, Australia \\
}

\pubyear{2015}
\volume{999}
\pagerange{998--999}
\date{?; revised ?; accepted ?}
\begin{document}

\maketitle

%
%
\begin{abstract}
We conducted direct numerical simulations (DNSs) of turbulent flow over three-dimen\-sional sinusoidal roughness in a channel. A passive scalar is present in the flow with Prandtl number $Pr=0.7$, to study heat transfer by forced convection over this rough surface.
The minimal-span channel is used to circumvent the high cost of simulating high Reynolds number flows, which  enables a range of rough surfaces to be efficiently simulated. 
The near-wall temperature profile in the minimal-span channel agrees well with that of the conventional full-span channel, indicating it can be readily used for heat-transfer studies at a much reduced cost compared to conventional DNS.
As the roughness Reynolds number, $k^+$, is increased, the Hama roughness function, $\Delta U^+$, increases in the transitionally rough regime before tending towards the fully rough asymptote of $\kappa_m^{-1}\log(k^+)+C$, where $C$ is a constant that depends on the particular roughness geometry and $\kappa_m\approx0.4$ is the von K{\'a}rm{\'an} constant. In this fully rough regime, the skin-friction coefficient is constant with bulk Reynolds number, $Re_b$. Meanwhile, the temperature difference between smooth- and rough-wall flows, $\Delta \Theta^+$, appears to  tend towards a constant value, $\Delta \Theta^+_{FR}$. This corresponds to the Stanton number (the temperature analogue of the skin-friction coefficient) monotonically decreasing with $Re_b$ in the fully rough regime.
Using shifted logarithmic velocity and temperature profiles, the heat transfer law as described by the Stanton number  in the fully rough regime can be derived once both the equivalent sand-grain roughness 
$k_s/k$ 
and the temperature difference $\Delta \Theta^+_{FR}$ are known.
In meteorology, this corresponds to the ratio of momentum and heat transfer roughness lengths, $z_{0m}/z_{0h}$, being linearly proportional to the inner-normalised momentum roughness length, $z_{0m}^+$, where the constant of proportionality is related to $\Delta \Theta_{FR}^+$.
While Reynolds analogy, or similarity between momentum and heat transfer, breaks down for the bulk skin-friction and heat-transfer coefficients, 
similar distribution patterns
between the heat flux and viscous component of the wall shear stress are observed.
Instantaneous visualisations of the temperature field show a thin thermal diffusive sublayer following the roughness geometry in the fully rough regime, resembling the viscous sublayer of a contorted smooth wall.
\end{abstract}


%
%
\section{Introduction}
\label{sect:intro}
Turbulent flow over rough surfaces is a problem inherent to many engineering and geophysical systems. Over the past few decades, especially with the advent of direct numerical simulation (DNS), we have seen substantial advances in the understanding of how roughness alters the overlying turbulent flow. In particular, the effect of various geometrical roughness parameters like height, wavelength and skewness on the drag force exerted on the wall has been quantified to a reasonably high level of accuracy \citep{Jimenez04,Flack14,Chan15}. However at present this advancement has not been mirrored in our depth of understanding of the effect of surface roughness on heat transfer. This is partly due to the separate communities researching roughness and heat transfer, as well as experimental difficulties in obtaining high fidelity temperature fields.
In this paper, we extend the minimal-span channel, recently shown to be capable of predicting the near-wall turbulent flow over roughness \citep{Chung15,MacDonald17}, to heat transfer. This significantly reduces the cost of the numerical simulations relative to conventional DNS while still retaining the same level of accuracy. 

Roughness generally increases the drag force exerted on the wall when compared to a smooth wall, which is often quantified by the (Hama) roughness function, $\Delta U^+$ \citep{Hama54}. This quantity reflects the retardation of the mean streamwise flow over a rough wall compared to a smooth wall. In the logarithmic region of the flow, the rough-wall velocity profile has the form
\begin{equation}
\label{eqn:logU}
U^+ = \frac{1}{\kappa_m}\log\left(z^+\right)+A_m-\Delta U^+,
\end{equation}
where $\kappa_m\approx0.4$ is the von K{\'a}rm{\'a}n constant, $A_m\approx5.0$ is the smooth-wall offset, and $z$ is the wall-normal position. The superscript $+$ indicates quantities non-dimensionalised on kinematic viscosity $\nu$ and friction velocity $U_\tau\equiv\sqrt{\tau_w/\rho}$, where $\tau_w$ is the wall-shear stress and $\rho$ is the fluid density.
The roughness function is related to the skin-friction coefficients, $C_f$, of smooth-wall (subscript $s$) and rough-wall (subscript $r$) flows at matched friction Reynolds numbers as \citep[e.g.][]{Schultz09}
\begin{equation}
\Delta U^+ = U_{bs}^+-U_{br}^+ = \sqrt{\frac{2}{C_{fs}}}-\sqrt{\frac{2}{C_{fr}}},
\end{equation}
where we are using the bulk velocity $U_b^+ =(1/h)\int_0^{h} U^+\mathrm{d}z$,  $h$ being the channel half-height, to define the skin-friction coefficient, $C_f\equiv \tau_w/(\tfrac{1}{2}\rho U_b^2)=2/U_b^{+2}$.
In the fully rough regime, in which the rough-wall skin-friction coefficient no longer depends on the bulk Reynolds number, $Re_b\equiv2hU_b/\nu$, the roughness function scales as $\Delta U^+=\kappa_m^{-1}\log(k^+)+C$, where $k$ is the roughness height and $C$ depends on the rough surface in question. 
If the offset $C$ is known, then extrapolations to arbitrary roughness Reynolds numbers, $k^+$, can be easily performed. Alternatively, the equivalent sand-grain roughness $k_s$ can be reported, which relates a given roughness length scale to the sand grain roughness size of \cite{Nikuradse33} as $k_s/k \equiv\exp[\kappa_m(3.5+C)]$. Here, the constant 3.5 comes from the difference between the smooth-wall log-law offset ($A_m\approx5$) and Nikuradse's rough-wall constant ($C_N\approx8.5$).
In the present DNS with three-dimensional sinusoidal roughness, we define the roughness height $k$ to be the sinusoidal semi-amplitude.

The temperature profile for forced convection smooth-wall flow also exhibits a region with logarithmic dependence on distance from the wall \citep{Kader81,Kawamura99,Pirozzoli16}. Roughness is generally a more efficient mechanism for heat transfer through the wall, which would result in a downwards shift of this logarithmic temperature profile \cite[e.g.][]{Yaglom79,Cebeci84}. Therefore, in the same manner as velocity, a temperature difference $\Delta \Theta^+$ can be defined as the difference in temperature between the smooth- and rough-wall flows in the outer layer of the flow. The rough-wall temperature profile would then follow
\begin{equation}
\label{eqn:logT}
\frac{\Theta-\Theta_w}{\Theta_\tau}=\frac{1}{\kappa_h}\log(z^+)+A_h(Pr)-\Delta \Theta^+,
\end{equation}
where the offset $A_h$ depends on the molecular Prandtl number, $Pr\equiv\nu/\alpha$, with $\alpha$ being the thermal diffusivity. The slope coefficient $\kappa_h\approx \kappa_m/Pr_t\approx0.46$ is slightly larger than that of the velocity profile in (\ref{eqn:logU}), owing to the turbulent Prandtl number, $Pr_t$, typically having a value less than unity, in the range of 0.85 to 0.9 \citep{Yaglom79,Pirozzoli16}. The temperature is relative to the wall temperature, $\Theta_w$, and is non-dimensionalised on the friction temperature, 
$\Theta_\tau\equiv (q_w/\rho c_p)/ U_\tau$, where $q_w$ is the  temporally and spatially averaged wall heat flux and $c_p$ is the specific heat at constant pressure. The validity of (\ref{eqn:logT}), particularly in the use of $\Delta \Theta^+$, has not received the same amount of attention as the roughness function, $\Delta U^+$. \cite{Miyake01} and \cite{Leonardi07tsfp} performed DNSs of two-dimensional spanwise-aligned square bars arranged on the bottom wall of a channel with a smooth top wall. Both studies prescribed different temperatures on the bottom and top walls and the resulting inner-normalised temperature profiles exhibited a downwards shift due to the roughness. \cite{Leonardi07tsfp} observed slightly different logarithmic slopes for varying aspect ratios of the bars, although this may be due to the asymmetric bottom and top walls. \cite{Chan15} discusses how this asymmetry can impact the perception of outer-layer similarity of the flow and may explain the different slopes for the temperature profiles. Moreover, \cite{Pirozzoli16} noted that channels with different prescribed temperatures between the bottom and top walls exhibit large wakes in the outer layer of the temperature profiles. This would hinder the identification of the logarithmic region, particularly with the relatively low friction Reynolds numbers that varied between the roughness cases used in \cite{Leonardi07tsfp}. In the present study, we will verify (\ref{eqn:logT}) using a scalar body-forcing approach with matched friction Reynolds numbers to address these uncertainties.

The heat transfer through the wall is often quantified by the Stanton number, $C_h=1/(U_b^+\Theta_m^+)$, which is the heat transfer analogue to the skin-friction coefficient. Here, we are using the mixed-mean (or cup-mixing) temperature,  $\Theta_{m} =\int_0^h U(\Theta-\Theta_w)\id z/\int_0^h U\id z$ \citep[e.g.][]{Owen63,Bird02,Pirozzoli16}.
This is the mean temperature that would result if the channel outlet discharged into a container and was well mixed.
It is different to the arithmetic mean temperature, $\Theta_a=(1/h)\int_0^h (\Theta-\Theta_w)\id z$, where
if logarithmic profiles are assumed for $U$ and $\Theta$, as in (\ref{eqn:logU}) and (\ref{eqn:logT}), this difference is
\begin{eqnarray}
\label{eqn:Tma}
\Theta_{m}^+ - \Theta_{a}^+ =  \frac{1}{\kappa_m\kappa_hU_b^+}.
\end{eqnarray}
Using this result and the definition of $C_h=1/(U_b^+\Theta_m^+)$, we see that the temperature difference, $\Delta \Theta^+$, between smooth- and rough-wall flows at matched friction Reynolds numbers  is related to the skin-friction and heat-transfer coefficients as
\begin{eqnarray}
\label{eqn:DT}
\Delta \Theta^+ &=& \Theta_{as}^+-\Theta_{ar}^+ \nonumber \\
&=& \left(\Theta_{ms}^+-\frac{1}{\kappa_m\kappa_h U_{bs}^+}\right) - \left(\Theta_{mr}^+-\frac{1}{\kappa_m\kappa_h U_{br}^+}\right) \nonumber \\
&=& \left(\frac{1}{U_{bs}^+C_{hs}}-\frac{1}{\kappa_m\kappa_h U_{bs}^+}\right) - \left(\frac{1}{U_{br}^+C_{hr}}-\frac{1}{\kappa_m\kappa_h U_{br}^+}\right) \nonumber \\
&=& \sqrt{\frac{C_{fs}}{2}}\left(\frac{1}{C_{hs}}-\frac{1}{\kappa_m\kappa_h}\right) - \sqrt{\frac{C_{fr}}{2}}\left(\frac{1}{C_{hr}}-\frac{1}{\kappa_m\kappa_h}\right).
\end{eqnarray}
Note that if the heat-transfer coefficient, $C_h$, was defined using the arithmetic mean temperature instead, then the temperature difference in (\ref{eqn:DT}) would not have the $1/{(\kappa_m\kappa_h)}$ terms. 
In external flows, such as zero-pressure-gradient boundary layers, the velocity and temperature scales to use are that of the freestream. This corresponds to the centreline velocity of internal flows, which are related to the bulk quantities through $U_h\equiv U(z=h)=U_b + 1/\kappa_m$ and $\Theta_h\equiv\Theta(z=h)=\Theta_a+1/\kappa_h$, where we have assumed logarithmic profiles across the entire channel, with no wake component.

The fully rough regime of roughness is associated with the pressure (or form) drag becoming dominant over the viscous drag \citep{Schultz09,Busse17}. The pressure drag component is independent of molecular viscosity, which is why in this regime the skin-friction coefficient is independent of the bulk Reynolds number. Heat transfer through the wall, however, always remains dependent on the molecular transport properties and there is no heat-transfer analogue to the pressure drag \citep{Owen63,Cebeci84}. Roughness increases the rate of heat transfer in the transitionally rough regime, however this increase cannot be sustained indefinitely as it is limited by the conductive heat flux near the surface.
In the fully rough regime this corresponds to the heat-transfer coefficient (Stanton number) monotonically decreasing with bulk Reynolds number, in a similar manner to the smooth wall but with an offset \citep{Owen63,Dipprey63}. However, this behaviour in terms of the temperature profile, in particular the behaviour of $\Delta \Theta^+$, has yet to be explained.

The present forced convection simulations neglect buoyancy effects, 
as the forces produced by the external source driving the flow 
(such as a pump producing a driving pressure gradient)
are typically much larger than  any buoyancy forces.
This is a common assumption made in engineering applications and smooth-wall DNS studies \citep[e.g.][]{Kim89scalar,Kasagi92,Kawamura98,Tiselj01,Pirozzoli16}.
Many geophysical flows, meanwhile, are driven by buoyancy, referred to as natural or free convection. However,
at very high Rayleigh numbers, sufficiently strong winds can develop that drive
 turbulent boundary layers near the wall that are characterised by local buoyancy forces that are much smaller than the shear forces.
 Studies of buoyancy-driven flows \citep[e.g.][]{Kraichnan62,Grossmann11,Ng17}
suggest that within this near-wall shear-dominated region the velocity and temperature profiles tend towards logarithmic functions of distance from the wall, similar to the
 present forced convection flow.
  This idea is encapsulated in Monin--Obukhov similarity theory, wherein buoyancy forces can be neglected when $|z/L|$ is small, where $L$ is the Obukhov length.
The present forced convection technique could  therefore be viewed as a tool to study the near-wall flow of high Rayleigh number natural convection systems, even though the global system is driven by buoyancy.

 Presently, we will describe the numerical procedure (\S\ref{sect:numerics}) and validate the minimal channel for forced convection heat transfer (\S\ref{sect:minchannel}). The roughness Reynolds number will then be increased from the transitionally rough regime towards the fully rough regime (\S\ref{sect:results}). This will enable us to investigate $\Delta \Theta^+$ and examine the resulting changes to the temperature field in the fully rough regime.

%
%
\section{Numerical procedure}
\label{sect:numerics}

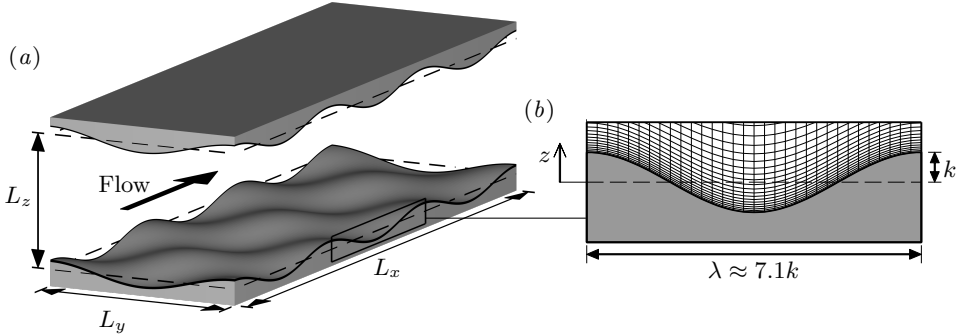
\begin{figure}
\centering
	\includegraphics[width=0.495\textwidth]{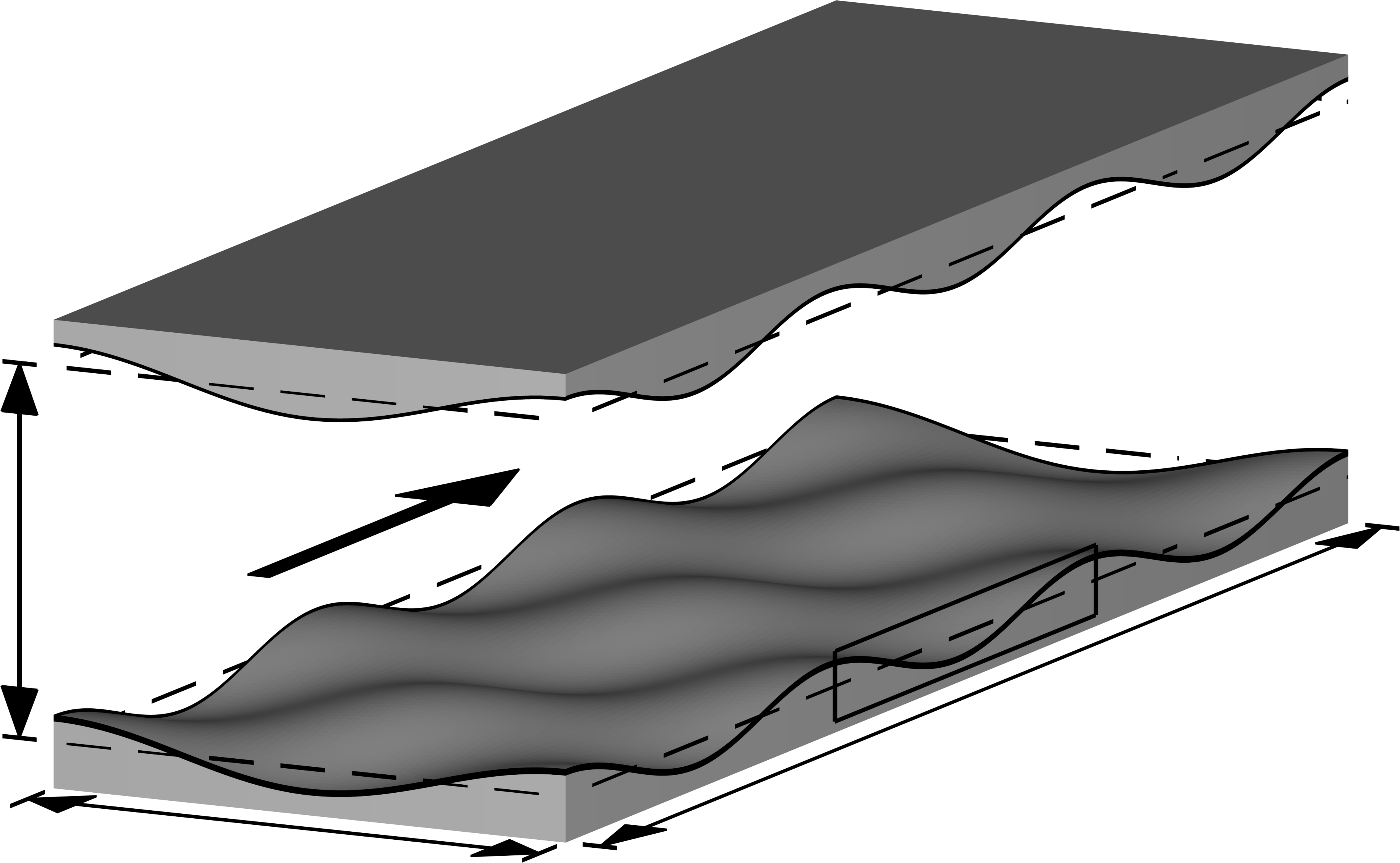}
	\includegraphics[width=0.4\textwidth,trim = -6 -25 0 0,clip = true]{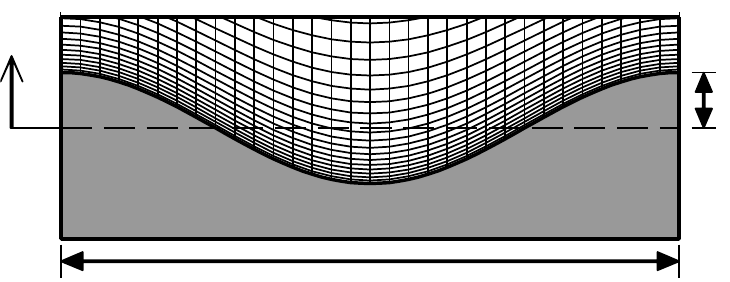}
	\put(-6.95,1.25){\line(1,0){2.13}}
	\put(-12.5,3.3){(\emph{a})}
	\put(-12.55,1.45){$L_z$}
	\put(-11.3,-0.2){$L_y$}
	\put(-7.65,0.5){$L_x$}
	\put(-11.3,1.6){Flow}
	\put(-5.65,2.6){(\emph{b})}
	\put(-3.2,0.45){$\lambda\approx7.1k$}
	\put(-0.07,1.8){$k$}
	\put(-5.45,2.0){$z$}
	\vspace{+0.0\baselineskip}
\caption{(\emph{a}) Channel domain and (\emph{b}) roughness geometry with semi-amplitude, $k$, and wavelength, $\lambda$.
Origin of $z$ is at the roughness mean height \citep{Chan15}. Only every fourth wall-normal node is shown in (\emph{b}).
}
	\label{fig:channelDomain}
\end{figure}

The numerical method used here is the second-order (kinetic) energy-conserving  finite volume code named CDP, designed for unstructured grids, with flow variables collocated at the cell centroids. Further details are available in \cite{Ham04} and \cite{Mahesh04}.  This is the same numerical method as has been used in our previous studies \citep{Chan15,MacDonald16}, although here we also solve the transport equation for a passive scalar representing temperature. It is assumed that there are no buoyancy effects as in forced convection, and we neglect temperature variations in viscosity. The conservation equations that are being solved are:
\begin{eqnarray}
\nabla\cdot \mathbf{u} &=& 0, \label{eqn:cont}\\
\pf{\mathbf{u}}{t}+\nabla\cdot(\mathbf{u}\mathbf{u}) &=& -\f{1}{\rho}\nabla p+ \nu\nabla^2 \mathbf{u}+\mathbf{F}, \label{eqn:mmtm}\\
\pf{\mathbf{\theta}}{t}+\nabla\cdot(\mathbf{u}\theta) &=& \alpha \nabla^2 \theta + G, \label{eqn:energy}
\end{eqnarray}
where $\mathbf{u}=(u,v,w)$ is the fluid velocity, 
$x$, $y$ and $z$ are the streamwise, spanwise and wall-normal (vertical) coordinates, respectively,
$t$ is time
and 
$\mathbf{F}=(F_x,0,0)$ is the driving pressure gradient term. In most turbulent channel flow studies, including our previous works, the pressure $p_T$ is decomposed into two components; the driving (mean) pressure, $P(x)$, and the fluctuating (periodic) pressure, $p$. The driving pressure $P$ is an input into the simulation through a spatially uniform body force $F_x = -(1/\rho)\dd P/\dd x>0$. This varies at each time step such that the bulk velocity is constant at all times. The pressure, that is solved for in (\ref{eqn:mmtm}) is the periodic component, $p$. 
A similar analogy can be made with temperature, in which the temperature $T_T$ is decomposed into the mean, $T(x)$ and periodic component, $\theta$, i.e. $T_T=T(x)+\theta$. The mean can be applied as a body force to the transport equation, $G = -u\cdot\dd T/\dd x$, where $u$ is the instantaneous, spatially dependent streamwise velocity, and the periodic component $\theta$ is solved for by the code. Physically, this amounts to a hot fluid (for $\dd T/\dd x<0$)  being cooled by the wall as it passes through the domain. For a statistically steady fully developed flow in which the time-averaged bulk temperature does not change with time, the heat added through this body forcing will be equal to the heat lost through the walls.
Here, the average heat flux at the wall is defined as $q_w/(\rho c_p)=\alpha \langle \overline{\partial \theta/\partial n}\rangle_w$, where $n$ is the local wall-normal direction, overline denotes temporal averaging and angled brackets denotes spatial averaging across all cell wall faces, $\langle \cdot \rangle_w\equiv (1/A_{p})\int_{wall}(\cdot)\id S$, such that $\langle 1 \rangle_w=A_w/A_p$ with $A_w$ being the wetted surface area and $A_p$ being the plan area.
An isothermal boundary condition for the walls is used with $\theta_w=0$, however the true wall temperature would be $T_w = T + \theta_w =T_0 + \dd T/\dd x\cdot x+0$, where $T_0$ is the reference temperature.  This is a non-conjugate heat transfer problem, or in other words the solid wall has an infinitely large thermal conductivity.  The present internal heating technique is similar to that employed in various isothermal smooth-wall heat transfer studies \citep{Kim89scalar,Kasagi92,Kawamura98,Tiselj01, Pirozzoli16}. The form that the scalar forcing term takes varies somewhat between these studies, which slightly alters the flow in the channel centre \citep{Pirozzoli16} but this should not affect the near-wall roughness effects that we are interested in.
 The Prandtl number is set to that of air at room temperature, $Pr=0.7$.  The bulk velocity for each case is set by trial and error such that the friction Reynolds number, $Re_\tau=U_\tau h/\nu$ is approximately equal to its target and  matched between smooth- and rough-wall flows. Here, $h$ is the channel half height, defined for the rough-wall flow to be distance between the channel centre and the roughness mean height \citep{Chan15}, corresponding to the hydraulic half height. Three-dimensional sinusoidal roughness is applied to both the bottom and top walls at $z=0+z_w$ and $z=2h+z_w$  (figure \ref{fig:channelDomain}), with
 \begin{equation}
 z_w=k\cos\left(\frac{2\pi x}{\lambda_x}\right)\cos\left(\frac{2\pi y}{\lambda_y}\right),
 \end{equation}
  where $k$ is the sinusoidal semi-amplitude and 
the sinusoidal wavelength  $\lambda_x=\lambda_y=\lambda\approx 7.07k$ matches that in our previous study \citep{Chung15}.
We use a body-fitted (terrain-following) grid to resolve the no-slip wall, as depicted in figure \ref{fig:channelDomain}(\emph{b}) and discussed in \cite{Chan15}.
The sinusoidal roughness has a wetted surface area approximately 17.8\% greater than the smooth wall.
 
\setlength{\unitlength}{1cm}

 Details of the simulations conducted are given in table \ref{tab:sims}. The minimal-span channel for rough wall flows is used \citep{Chung15,MacDonald17}, in which the spanwise domain width is very narrow and only the near-wall flow is captured up to a critical height $z_c^+\approx0.4L_y^+$, where $L_y^+$ is the channel span.  The recommendation in \cite{Chung15} is used to determine the spanwise domain width, namely $L_y\gtrsim \max(100\nu/U_\tau,k/0.4,\lambda$), where $\lambda$ is the spanwise roughness length scale. 
 The streamwise length should satisfy $L_x\gtrsim\max(3L_y,1000\nu/U_\tau,\lambda_{r,x})$, as discussed in \cite{MacDonald17}.  
Simulations are run for between $120$ to $600$ large-eddy turnover times $z_c/U_\tau$ (depending on $L_y^+$ and $Re_\tau$) to ensure that the uncertainty in $\Delta U^+$ is less than 0.1$U_\tau$, following the guidelines in \cite{MacDonald17}.
 Smooth-wall channel simulations with matched channel domain sizes have also been conducted, to ensure that the differences between the smooth- and rough-wall flows are due to the roughness alone and not the channel span. In set $A$, we simulate spans of $L_y^+=155=\lambda^+$, $L_y^+=310=2\lambda^+$ and a full-span channel with $L_y=\pi h$ at $Re_\tau=395$ to assess the impact of this spanwise width on the heat transfer. 
In set $B$, we simulate two different friction Reynolds numbers of $Re_\tau=395$ and $Re_\tau=590$ but with matched roughness viscous dimensions (same $k^+$ and $\lambda^+$) and channel viscous dimensions, to examine the effect of relatively low $Re_\tau$ on the flow.
 Finally, in set $C$, we then increase the
 roughness Reynolds number, $k^+$, towards the fully rough regime. In this set, all cases have $k=h/18$ except for the first case where $k=h/36$ (to ensure that $Re_\tau\gtrsim395$).
  The expected full-span bulk velocity, $U_{bf}^+=\int_0^{h} U_{f}^+\mathrm{d}z/h$, is given in table \ref{tab:sims}, where the expected full-span velocity profile $U_f$ is defined such that the simulation data from the minimal channel is used for $z<z_c$, while the composite velocity profile of \cite{Nagib08} for full-span channel flow is used  for $z>z_c$. The log-law offset constant is set such that $U_f$ is continuous at $z=z_c$. 
  A hyperbolic tangent grid stretching is used in the wall-normal (vertical) direction, resulting in a fairly large grid spacing at the channel centreline. However, the grid spacings below $z_c$ are  such that $\Delta z^+$ only increases beyond conventional DNS spacings above the vertical critical height, $z_c$. As the region  of the flow above $z_c$ is already altered due to the nature of the minimal channel, these spacings should have negligible impact on the near-wall flow of interest. 
  A uniform grid spacing is used in the streamwise and spanwise directions. Horizontal (wall-parallel) averaging is performed using the intrinsic spatial average for $z<k$, in which quantities represent averages over only the fluid regions. This can be related to the superficial spatial average (averaging over both fluid and solid regions) by multiplying the intrinsic average by the ratio of fluid to total volume, $\sigma(z)$ \citep[see e.g.][]{Finnigan00,Nikora01,Breugem06}.

\begin{table}
\centering
\begin{tabular}{p{0.35cm} c c c c c c c c c c c c c c}
& $Re_\tau$	& $\frac{h}{k}$	& $k^+$	& $\lambda^+$	& $L_x^+$	& $L_y^+$	& $N_x$	& $N_y$	& $N_z$	& $\Delta z_w^+$	& $\Delta z_h^+$	& $U_{bf}^+$	& $\Delta U^+$	& $\Delta \Theta^+$\\[+0.2em]
\multirow{2}{*}{${A}\left\{ \rule{0cm}{0.6cm}\right.$}
&395			& 18			& 21.9	& 155		& 1086	& 155	& 224	& 32		& 320	& 0.35			& 6.9				& 10.5		& 6.5 		& 3.0 \\
&395			& 18			& 21.9	& 155		& 1086	& 310	& 224	& 64		& 320	& 0.35			& 6.9 			& 10.4		& 6.8 		& 3.1 \\
&395			& 18			& 21.9	& 155		& 2482	& 1241	& 512	& 256	& 320	& 0.35			& 6.9 			& 10.5		& 6.7 		& 2.9 \\[+0.4em]

\multirow{1}{*}{${B}\left\{ \rule{0cm}{0.39cm}\right.$}
&395			& 18			& 21.9	& 155		& 1086	& 155	& 224	& 32		& 320	& 0.35			& 6.9				& 10.5		& 6.5 		& 3.0 \\
&590			& 27			& 21.9	& 155		& 1086	& 155	& 224	& 32		& 380	& 0.29			& 7.3 			& 11.6		& 6.5 		& 3.1 \\[+0.4em]

\multirow{5}{*}{${C}\left\{ \rule{0cm}{1.2cm}\right.$}
&395			& 36			& 11.0	& 78		& 1086	& 155	& 448	& 64		& 320	& 0.28			& 5.5 			& 13.1		& 4.0 		& 1.8 \\
&395			& 18			& 21.9	& 155		& 1086	& 155	& 224	& 32		& 320	& 0.35			& 6.9 			& 10.5		& 6.5 		& 3.0 \\
&590			& 18			& 32.8	& 232		& 1390	& 232	& 288	& 48		& 380	& 0.29			& 7.3 			& 9.9			& 8.2 		& 3.9 \\
&720			& 18			& 40.0	& 282		& 1414	& 282	& 320	& 64		& 480	& 0.22			& 7.4 			& 9.7			& 9.0 		& 4.2 \\
&1200			& 18			& 66.7	& 471		& 1414	& 471	& 306	& 102	& 720	& 0.25				& 8.3 			& 9.7			& 10.3 	& 4.5 \\
&1680			& 18			& 93.3	& 660		& 1979	& 660	& 432	& 144	& 800	& 0.31				& 10.4 		& 9.7			& 10.9	& 4.3 \\
\end{tabular}
\vspace{-0.3\baselineskip}
\caption{Description of the different rough-wall simulations performed.  For each roughness case, a smooth-wall case with matched channel dimensions is also simulated.
$L_x^+$ and $L_y^+$ are the channel length and width;
$N_x$, $N_y$ and $N_z$ are the number of cells in the streamwise, spanwise and wall-normal (vertical) direction;
$\Delta z_w^+$ and $\Delta z_h^+$ are the wall-normal grid spacings at the wall and channel centre, respectively.
$U_{bf}^+$ is the expected full-span bulk velocity using a composite velocity profile
and $\Delta U^+$ ($\Delta \Theta ^+$) is the roughness function (temperature difference) computed from the difference in smooth- and rough-wall velocities (temperatures) evaluated at $z_c=0.4L_y$.
}
\label{tab:sims}
\end{table}

%
%
\section{Heat transfer in the minimal channel}
\label{sect:minchannel}
\subsection{Effect of channel width, $L_y^+$ (set $A$)}
\label{ssect:minchannelLy}

\setlength{\unitlength}{1cm}
\begin{figure}
\centering
	\includegraphics[width=0.495\textwidth]{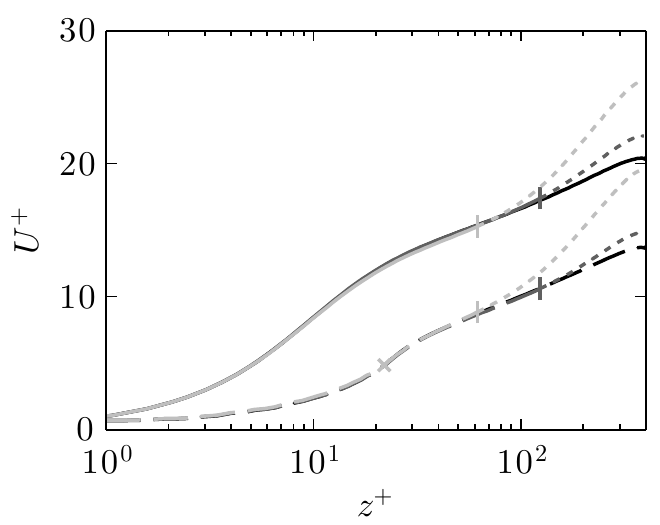}
	\includegraphics[width=0.495\textwidth]{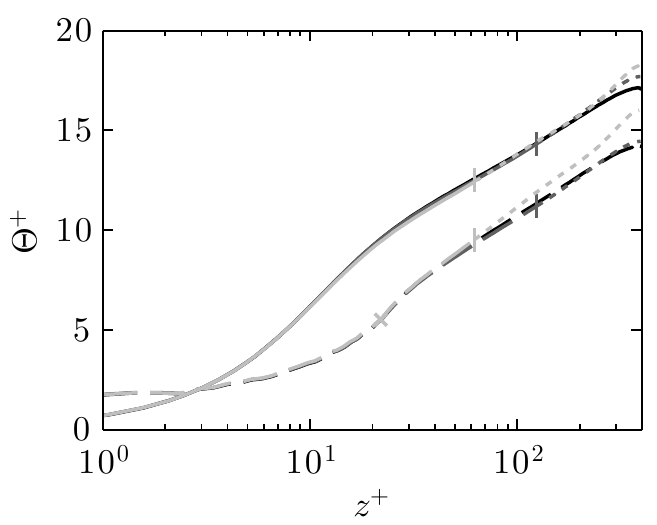}
	\put(-12.3,2.6){\includegraphics[width=0.22\textwidth]{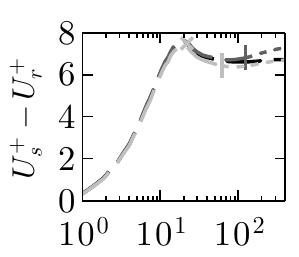}}
	\put(-5.6,2.6){\includegraphics[width=0.22\textwidth]{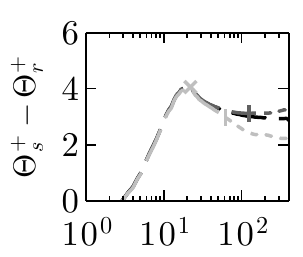}}
	\put(-13.5,5.0){(\emph{a})}
	\put(-6.65,5.0){(\emph{b})}
	\vspace{-0.5\baselineskip}
\caption{Mean (\emph{a}) velocity  and (\emph{b})  temperature profiles for smooth-wall (solid) and rough-wall (dashed) channels with $k^+\approx 22$ (cross).
Darker grey refers to increasing channel width (set $A$, table \ref{tab:sims}). 
Dotted lines indicate unphysical region above the wall-normal critical height $z_c=0.4L_y$ (vertical mark).
Insets show difference in smooth- and rough-wall profiles.
}
	\label{fig:veltempWidth}
\end{figure}

First, we will consider the effect of the channel width on mean velocity and temperature profiles, for both smooth- and rough-wall flows. 
This is done at a matched friction Reynolds number of $Re_\tau\approx395$ (set $A$, table \ref{tab:sims}). 
The rough-wall minimal channel has been previously validated for momentum transfer, where it has been shown to be capable of reproducing the roughness function as well as the near-wall high-order statistics of a conventional full-span channel \citep{Chung15,MacDonald16, MacDonald17}.
In the present forced convection flow, temperature is a passive scalar and is simply advected by the velocity, suggesting that it will respond in the same manner as velocity to the minimal-span channel. However, \cite{Pirozzoli16} noted some differences between the passive scalar and velocity fields, especially in the outer (core) region of the flow. We will therefore compare mean velocity and temperature profiles for minimal and full-span channels.
The mean streamwise velocity profile is shown in figure \ref{fig:veltempWidth}(\emph{a}), where the dotted lines indicate the unphysical region above the wall-normal critical height, $z_c=0.4L_y$, of the minimal channels. As expected, this scaling agrees well with the data, where we see that above the critical height (denoted by the vertical mark) the mean velocity increases relative to the full-span channel. Below this point we have what can be described as `healthy' turbulence \citep{Flores10}, as it is the same as in conventional (full-span) channel flow. Widening the channel by increasing $L_y^+$ extends the region of healthy turbulence further from the wall, as larger turbulent structures can fit inside the widened domain. 
The region above $z_c$ is unphysical due to the narrow constraints of the channel and is not relevant to the near-wall flow.
 The inset of this figure shows the difference in smooth- and rough-wall velocity profiles as a function of $z^+$. This difference reaches a constant above $z^+\approx 40$, which is the offset $\Delta U^+$ of the rough-wall flow relative to the smooth-wall flow. Since the difference is constant with $z^+$, this indicates that the rough-wall flow has the same velocity profile as the smooth-wall flow and that the outer-layers are similar. This is the case for both minimal-span and full-span channels and demonstrates that the minimal channel can accurately estimate the roughness function, as already discussed in our previous work \citep{Chung15,MacDonald17}.

Figure \ref{fig:veltempWidth}(\emph{b}) shows the temperature profile in the same format as the velocity profile of figure \ref{fig:veltempWidth}(\emph{a}). The critical height scaling obtained from the velocity profiles, $z_c=0.4L_y$, is used here without alteration. There is excellent near-wall agreement between the minimal and conventional channels and it appears that the mean temperature does not increase as readily as the velocity when we are outside the healthy turbulence region (above $z_c$, dotted lines). As with the velocity profiles, both the smooth- and rough-wall temperature profiles appear to scale with logarithmic distance from the wall above $z^+\gtrsim50$.
The inset shows the difference in smooth- and rough-wall temperature profiles,  which for the conventional full-span channel flow (black line) is tending towards a constant value value above $z^+\approx60$. As with the velocity profile, this indicates that the smooth- and rough-wall temperature profiles are similar in the outer layer of the flow. 
This supports the rough-wall logarithmic temperature profile (\ref{eqn:logT}), and that we only need to estimate the temperature equivalent of the roughness function, $\Delta \Theta^+$, to describe the temperature profile.
 The difference in temperature profiles for the minimal-span channels  (grey lines in inset of figure \ref{fig:veltempWidth}\emph{b}) shows more variation above $z_c^+$ than the corresponding velocity difference. The narrowest span channel with $z_c^+\approx 62$ (light grey line) shows the temperature difference tending towards a value of approximately 2.2 at the channel centreline, much lower than the conventional channel centreline value (black line) of 3.0.  This difference in the minimal-span channel is possibly due to a lack of statistical convergence in the outer-layer region of the channel and would require a much longer run time to converge. However, the region above $z_c^+$ is inherently unphysical due to the minimal span and resulting lack of large-scale structures, which means obtaining statistical convergence in this region is not necessary or relevant to the near-wall healthy turbulence \citep{MacDonald17}. The minimal channel simulations are therefore run to ensure converged statistics in the near-wall flow, up to the critical height, $z_c^+$. As such, if we evaluate the temperature difference at the critical height $z_c^+$ to obtain $\Delta \Theta^+$,we observe good agreement between all three channel widths, with $\Delta \Theta^+\approx $ 3.0, 3.1 and 2.9 for $L_y^+=155$, 310 and the full-span case, respectively.
This result indicates that the minimal channel can be used for studying heat transfer, for both smooth- and rough-wall flows. 

\subsection{Effect of friction Reynolds number, $Re_\tau$ (set $B$)}
\label{ssect:retau}

\setlength{\unitlength}{1cm}
\begin{figure}
\centering
	\includegraphics[width=0.495\textwidth]{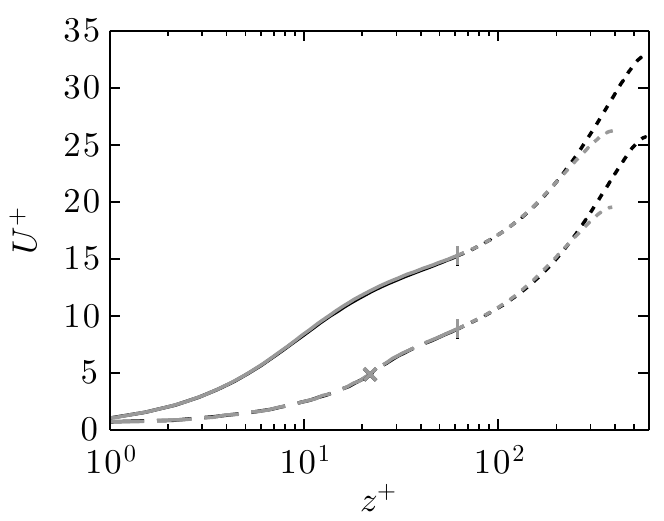}
	\includegraphics[width=0.495\textwidth]{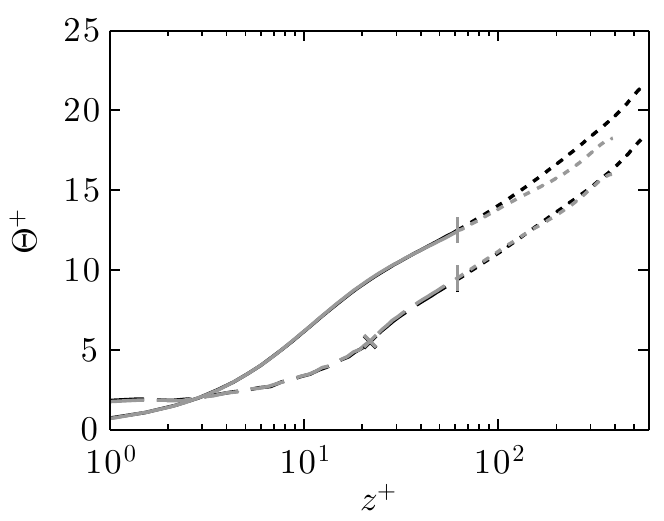}
	\put(-12.1,2.5){\includegraphics[width=0.22\textwidth]{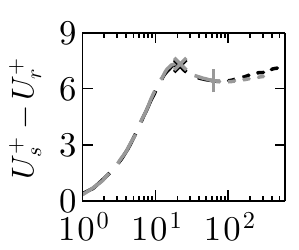}}
	\put(-5.4,2.5){\includegraphics[width=0.22\textwidth]{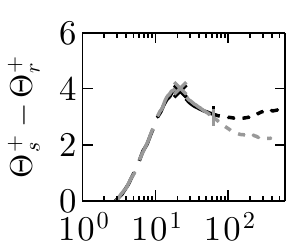}}
	\put(-13.5,5.0){(\emph{a})}
	\put(-6.65,5.0){(\emph{b})}
	\vspace{-0.5\baselineskip}
\caption{
Mean (\emph{a}) velocity and (\emph{b}) temperature profiles for smooth-wall (solid) and rough-wall (dashed) channels with $k^+\approx 22$ (cross), at $Re_\tau\approx 395$ (grey) and $Re_\tau\approx590$ (black) (set $B$, table \ref{tab:sims}). 
Dotted lines indicate unphysical region above the critical height $z_c=0.4L_y$ (vertical mark).
Insets show difference in smooth- and rough-wall profiles.
}
	\label{fig:veltempRetau}
\end{figure}

Having validated that the minimal channel can accurately predict both the roughness function, $\Delta U^+$, and the temperature difference, $\Delta \Theta^+$, we now assess the influence of the friction Reynolds number, $Re_\tau$. In \cite{Chan15}, it was shown that turbulent pipe flow simulations at $Re_\tau\approx180$ leads to an overestimation of the roughness function compared to $Re_\tau\gtrsim360$. This is primarily due to an upward shift of the logarithmic region in the smooth-wall case, caused by the pressure gradient effect inherent to low $Re_\tau$ turbulent flows. Here, we repeat this validation but also investigate the effect on heat transfer for two cases of $Re_\tau=395$ and $Re_\tau=590$ (set $B$, table \ref{tab:sims}). The roughness size ($k^+$ and $\lambda^+$) and channel dimensions ($L_x^+$ and $L_y^+$) are matched in viscous units. Figure \ref{fig:veltempRetau} shows the mean velocity and temperature profiles for these two friction Reynolds numbers. As already demonstrated in \cite{Chan15}, the mean velocity profiles (figure \ref{fig:veltempRetau}\emph{a}) and velocity difference (inset) shows that the Reynolds number is effect is negligible for $Re_\tau\gtrsim395$. The roughness function is marginally overestimated for $Re_\tau\approx395$ by about 0.08$U_\tau$, although this is similar to the level of uncertainty we would expect from the minimal-span channel \citep{MacDonald17}.

The temperature profiles (figure \ref{fig:veltempRetau}\emph{b}) and difference (inset) shows a similar effect, with there being only a minor difference between $Re_\tau\approx395$ and $Re_\tau\approx590$. Much of this difference is in the outer-layer region of the smooth-wall flow at $Re_\tau\approx395$ (solid grey line). However the temperature difference (inset) is similar below the critical height $z_c^+$, with the $Re_\tau\approx395$ case underestimating $\Delta \Theta^+$ by approximately $0.1\Theta_\tau$. While this result does not give a lower bound on the friction Reynolds number for which Reynolds number effects become negligible in heat transfer, we will assume that, as with velocity, $Re_\tau\gtrsim395$ is sufficient.


\section{Increasing roughness Reynolds number (set $C$)}
\label{sect:results}
\subsection{Mean profiles}
\label{ssect:profiles}

\setlength{\unitlength}{1cm}
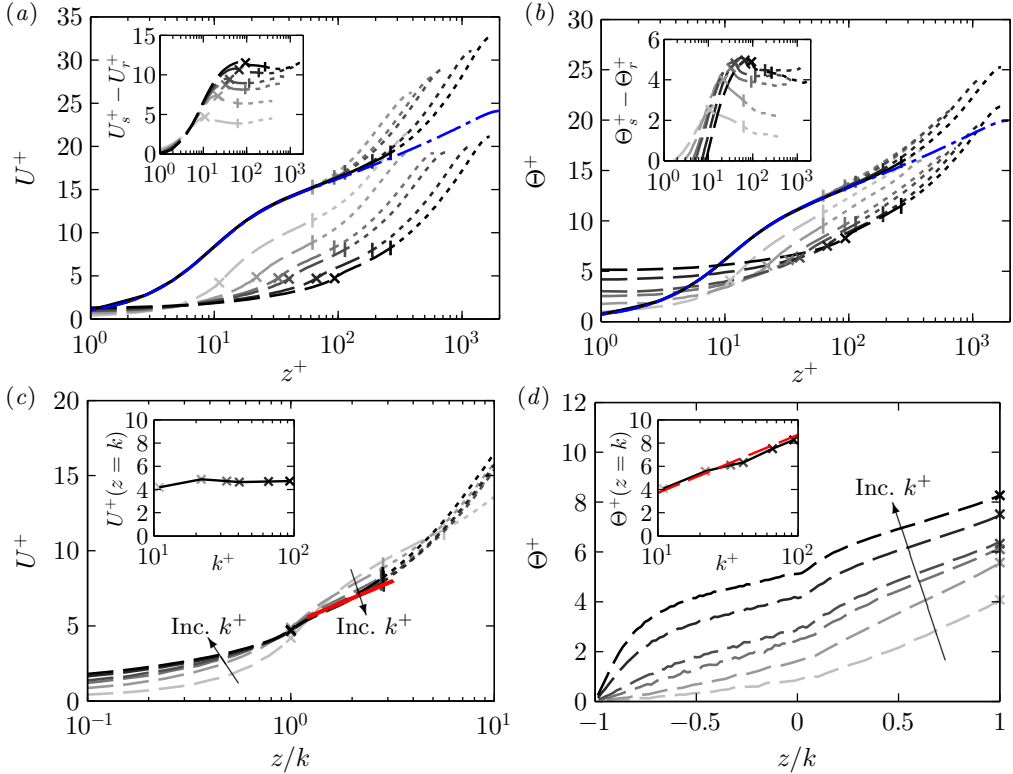
\begin{figure}
\centering
	\includegraphics[width=0.495\textwidth]{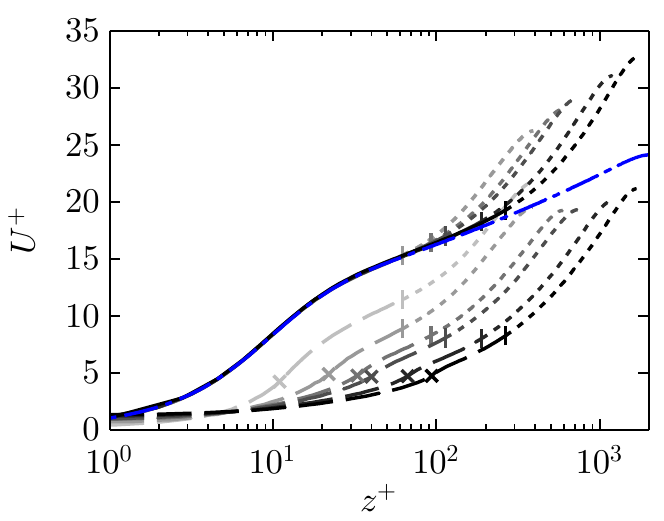}
	\includegraphics[width=0.495\textwidth]{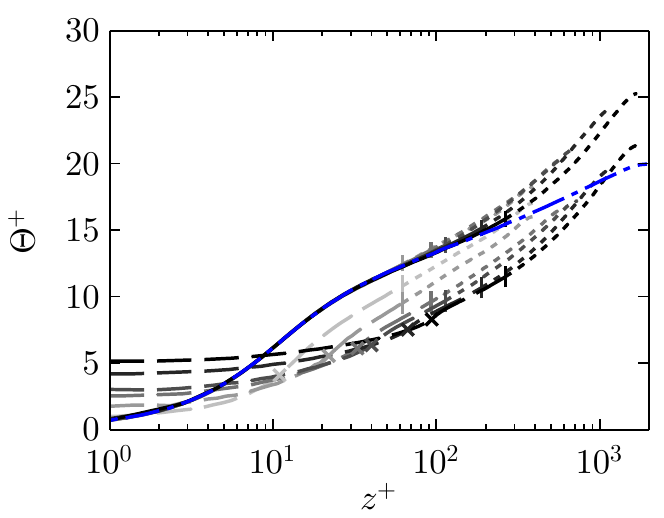}
	\put(-12.25,2.65){\includegraphics[width=0.22\textwidth]{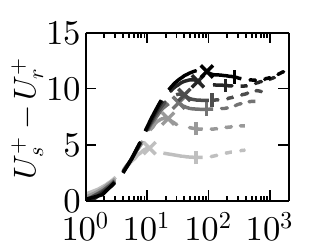}}
	\put(-5.5,2.55){\includegraphics[width=0.22\textwidth]{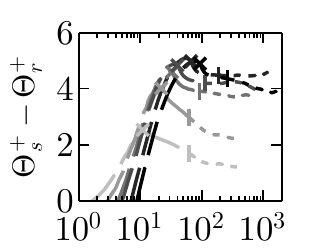}}
	\put(-13.5,4.9){(\emph{a})}
	\put(-6.6,4.9){(\emph{b})}
	\\
	\vspace{-0.5\baselineskip}
	\includegraphics[width=0.495\textwidth,trim=0 0 7 0, clip = true]{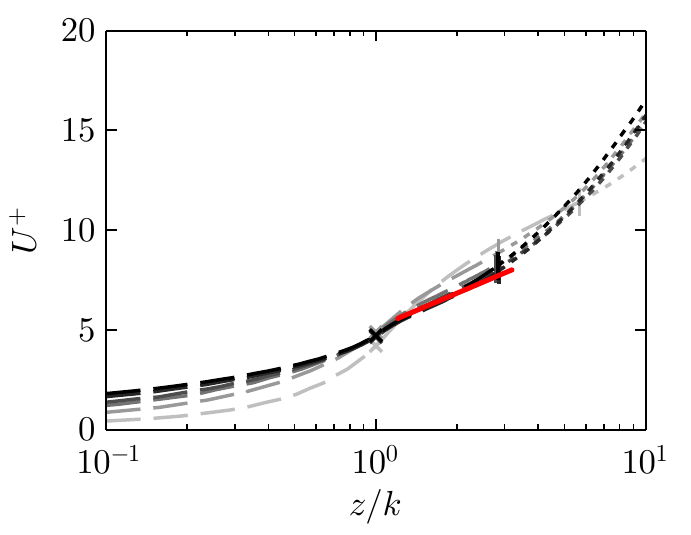}
	\includegraphics[width=0.495\textwidth]{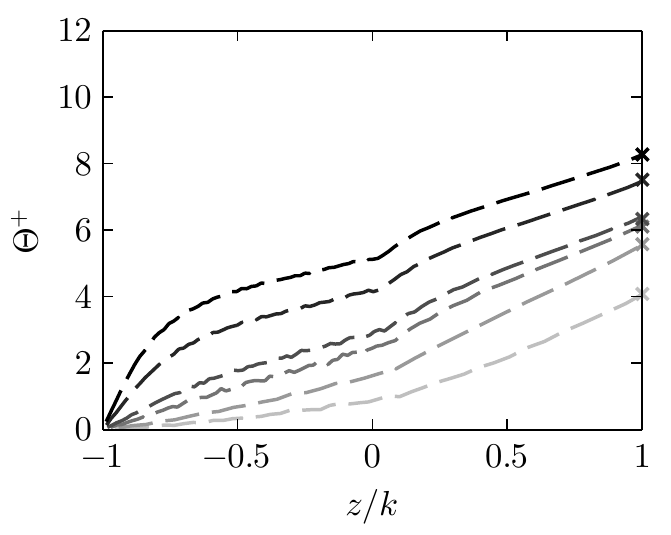}
	\put(-12.25,2.7){\includegraphics[width=0.22\textwidth]{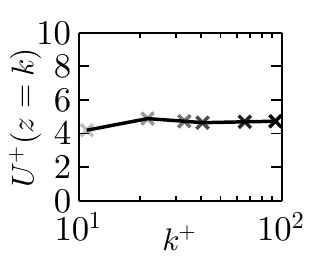}}
	\put(-5.55,2.7){\includegraphics[width=0.22\textwidth]{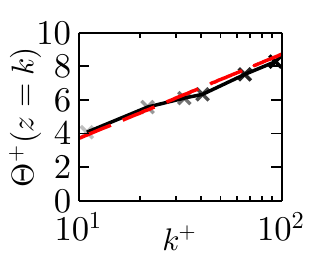}}
	\put(-10.4,1.3){\vector(-2,3){0.4}}
	\put(-11.3,1.95){Inc.~$k^+$}
	\put(-8.9,2.8){\vector(1,-3){0.2}}
	\put(-9.1,1.95){Inc.~$k^+$}
	\put(-13.5,5.0){(\emph{c})}
	\put(-6.7,5.0){(\emph{d})}
	\put(-1,1.6){\vector(-1,3){0.70}}
	\put(-2.2,3.8){Inc.~$k^+$}
	\vspace{-0.5\baselineskip}
\caption{(Colour online) Mean (\emph{a}) velocity  and (\emph{b}) temperature profiles for smooth-wall (solid) and rough-wall (dashed) channels with increasing friction Reynolds numbers.
Darker grey refers to increasing roughness height (set $C$, table \ref{tab:sims}). 
Insets show difference in smooth- and rough-wall profiles.
Dotted lines indicate unphysical region above the critical height $z_c=0.4L_y$  (vertical mark).
Blue dash-dotted line in (\emph{a}) and (\emph{b}) is the smooth-wall DNS data of \cite{Bernardini14} and \cite{Pirozzoli16}, respectively. 
(\emph{c}) Rough-wall velocity profiles against wall-normal position normalised on roughness height, $z/k$. Solid red line is the fully rough asymptote, $U^+\approx(1/0.4)\log(z/k) + 5.1$.
(\emph{d}) Rough-wall temperature profiles below the roughness crest. 
Insets in (\emph{c}) and (\emph{d}) show the velocity and temperature at the roughness crest, against $k^+$.
Dashed red line in inset of (\emph{d}) is the logarithmic temperature profile (\ref{eqn:logT}) evaluated at $z=k$.
}
	\label{fig:veltempIncK}
\end{figure}

We now increase the roughness Reynolds number, $k^+$, towards fully rough conditions (set $C$, table \ref{tab:sims}). Figure \ref{fig:veltempIncK}(\emph{a}) shows the mean velocity profiles for increasing friction Reynolds numbers, where the roughness height is fixed at $k=h/18$
(except for the smallest roughness case where $k=h/36$, to ensure $Re_\tau\gtrsim395$, see \S\ref{ssect:retau}).
 We see that increasing $Re_\tau$ (and hence $L_y^+=\lambda^+\approx7.1k^+$) leads to the capturing of a larger region of the logarithmic layer for the smooth-wall flow (solid lines). However, the roughness increasingly reduces the near-wall velocity magnitude with Reynolds number (dashed lines), indicating that the roughness function is increasing (inset of figure \ref{fig:veltempIncK}\emph{a}). The rough-wall velocity profiles are plotted in figure \ref{fig:veltempIncK}(\emph{c}) as a function of $z/k$. In this scaling, we see that the cases with larger $k^+$ (darker grey lines) are collapsing onto the fully rough asymptote of $U^+\approx \kappa_m^{-1}\log(z/k)+D$ (solid red line), where $\kappa_m\approx0.4$ and $D=A_m-C\approx 5.1$.
 In a conventional flow we would also see the centreline velocity of the rough-wall flow tending towards a constant as the skin-friction coefficient becomes constant in the fully rough regime. This is not observed here as the channel width, $L_y=\lambda$, increases with Reynolds number which affects the critical height, $z_c^+$, and hence centreline velocity. However, the velocity at the roughness crest (inset of figure \ref{fig:veltempIncK}\emph{c}) is approximately constant, $U_k^+\equiv U^+(z=k)\approx 4.7$, indicating that the drag coefficient defined on this velocity \cite[e.g.,][]{Macdonald98,Coceal04} is also approximately constant with Reynolds number, $C_{dk}\equiv\tau_w/((1/2)\rho U_k^2)=2/{U_k^{+2}}\approx0.09$.

Figure \ref{fig:veltempIncK}(\emph{b}) shows the temperature profile for the smooth- and rough-wall flows. The smooth-wall DNS data of \cite{Pirozzoli16} at $Re_\tau\approx2000$ is also shown by the dash-dotted blue line. We see excellent agreement with the present smooth-wall data below the critical height $z_c^+$ (vertical marks), further supporting the view made in \S\ref{ssect:minchannelLy} that the minimal channel can be used to study the near-wall flow of forced convection heat transfer.
\cite{Pirozzoli16} determined that the smooth-wall temperature profile in the logarithmic region follows $\Theta^+=\kappa_h^{-1}\log(z^+)+A_h$, with constants $\kappa_h\approx0.46$ and $A_h(Pr=0.7)\approx 3.2$. As the roughness Reynolds number increases, the rough-wall temperature profiles (dashed lines) begin to collapse and follow a logarithmic trend with $z^+$, the same as with the smooth-wall (solid lines) but offset. This agrees with the rough-wall logarithmic temperature profile introduced through (\ref{eqn:logT}).
The temperature difference (inset of figure \ref{fig:veltempIncK}\emph{b}) begins to collapse as well for large $k^+$, indicating that $\Delta \Theta^+$ is reaching a constant of approximately 4.4 that is independent of $k^+$. 
While the temperature profile is collapsing in the logarithmic region for fully rough flow (figure \ref{fig:veltempIncK}\emph{b}), the temperature below the roughness crests continues to increase with $k^+$, shown in figure \ref{fig:veltempIncK}(\emph{d}). 
This is consistent with the logarithmic temperature profile (\ref{eqn:logT}), which becomes independent of $k^+$ in the fully rough regime, beginning almost immediately above the crest. This would then force the crest temperature to be $\Theta_k^+\equiv\Theta^+(z=k)\approx\kappa_h^{-1}\log(k^+)+A_h-\Delta \Theta^+$, with $A_h-\Delta \Theta^+\approx3.2-4.4\approx-1.2$. Indeed, the inset of this figure shows that the temperature at the crest, $\Theta_k^+$, increases logarithmically with $k^+$, with good agreement with the logarithmic temperature profile (\ref{eqn:logT}) evaluated at $z=k$ (dashed red line).

\setlength{\unitlength}{1cm}
\begin{figure}
\centering
	\includegraphics{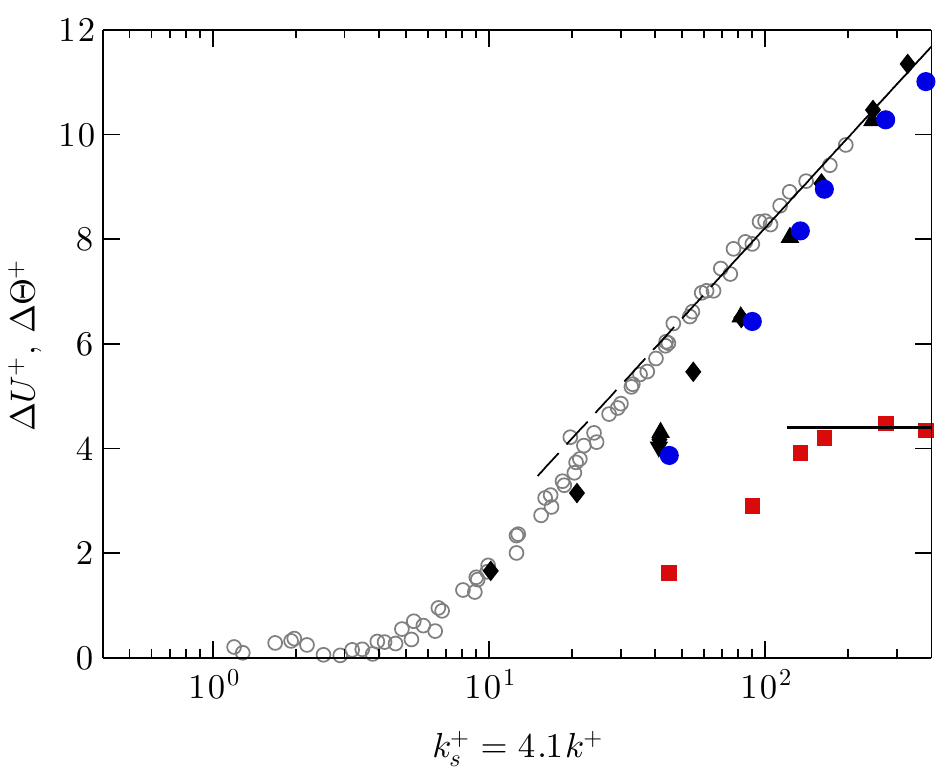}
	\put(-6.8,6.3){$\Delta U^+_{FR} = (1/\kappa_m)\log(k_s^+)+A_m-8.5$}
	\put(-3.5,6.1){\vector(1,-3){0.57}}
	\put(-2.05,4.0){$\Delta \Theta^+_{FR} \approx 4.4$}
	\put(-1.3,3.9){\vector(0,-1){0.37}}
	\vspace{-0.5\baselineskip}
\caption{
(Colour online)
Roughness function, $\Delta U^+$ (\protect\raisebox{-0.5ex}{\protect\scalebox{2.0}{$\color{myblue}{\bullet }$}})
and
temperature difference, $\Delta \Theta^+$ (\protect\raisebox{-0.1ex}{\protect\scalebox{1.0}{$\color{myred}\blacksquare$}}) 
for the present sinusoidal roughness, as a function of the equivalent sand-grain roughness, $k_s/k\approx 4.1$. 
Other symbols are the roughness function for
the sand grain data of \cite{Nikuradse33} (\protect\raisebox{0.2ex}{\protect\scalebox{0.8}{$\color{LGrey}\bigcirc$}})
and for
the same sinusoidal roughness geometry of \cite{Chung15} and \cite{Chan15} 
in
a minimal channel ($\color{black}\blacktriangle$),
a full channel ($\color{black}\blacktriangledown$),
and
a pipe ($\color{black}\blacklozenge$).
}
	\label{fig:DUDT}
\end{figure}

Figure \ref{fig:DUDT} shows the roughness function, $\Delta U^+$, and the temperature difference, $\Delta \Theta^+$, for the present roughness simulations.
This is shown with the roughness functions from \cite{Chung15} and \cite{Chan15} for the same  sinusoidal roughness geometry, as well as for the sand grain data of \cite{Nikuradse33}.  
The sinusoidal roughness data are plotted as a function of the equivalent sand-grain roughness, $k_s/k\approx4.1$, where this factor has been taken from \cite{Chung15}.
This scaling ensures the collapse of the present $\Delta U^+$ values with those of Nikuradse's sand grain roughness in the fully rough regime (here, $k_s^+\gtrsim150$), 
where the roughness function scales as 
$\kappa_m^{-1}\log(k_s^+)+A_m-C_N\approx\kappa_m^{-1}\log(k_s^+)-3.5$.
Within the transitionally rough regime ($k_s^+\lesssim150$), the different roughness geometries have a unique behaviour and the roughness function is not guaranteed to collapse with that of Nikuradse \citep{Jimenez04}.
Despite having matched roughness geometries, there is a slight difference between the pipe (black diamonds) and present minimal channel  (blue circles) data for the largest $k_s^+$. This is likely
 due to the fundamental differences between the two domain geometries, as well as the difference in blockage ratios ($k/h=1/6.75$ from \cite{Chan15} versus $k/h=1/18$ here).

From the temperature difference in figure \ref{fig:DUDT} it appears that, for small $k_s^+$, there may be a `thermodynamically smooth' regime for heat transfer in which $\Delta \Theta^+\approx 0$. This would be analogous to the hydrodynamically smooth regime for momentum transfer, in which $\Delta U^+\approx 0$ for $k_s^+\lesssim 4$  and the drag produced by the rough wall matches that of the smooth wall \citep{Raupach91,Jimenez04}. A visual extrapolation of the present data to small $k_s^+$ suggests that this thermodynamically smooth regime would remain at larger $k_s^+$ values than the hydrodynamically smooth regime.
This implies that for small $k_s^+$ the effects of roughness are first felt in momentum transfer before affecting heat transfer,
presumably due to the present molecular Prandtl number being less than unity ($Pr=0.7$).
In this case, the thermal diffusive sublayer is slightly thicker than the viscous sublayer and would require a larger $k_s^+$ to overcome.

In the fully rough regime, the temperature difference appears to tending towards a constant value of $\Delta \Theta^+\approx 4.4$. This implies that further increases to the roughness Reynolds number result in the heat-transfer coefficient (Stanton number) decreasing, consistent with the experiments of \cite{Dipprey63}. Presumably this constant $\Delta \Theta_{FR}^+$ would be affected by the roughness geometry, and would depend on the solidity (a measure of the roughness density) among other roughness parameters.
Note that when viewed in isolation, it may appear that $\Delta \Theta^+$ could have reached a maximum around $k_s^+\approx 250$ and could go on to decrease at larger Reynolds numbers, as opposed to remaining constant.  However, the flow appears to have reached its asymptotic fully rough state, where $\Delta U^+$ scales as $(1/\kappa_m)\log(k_s^+)$ and $U_k^+$ is constant (inset of figure \ref{fig:veltempIncK}\emph{c}). It is therefore difficult to conceive how the flow could undergo an additional change at even higher Reynolds numbers, well within the fully rough regime, that would lead to $\Delta \Theta^+$ decreasing. This decrease would imply that the temperature profile would eventually return to that of the smooth-wall flow while the smooth- and rough-wall velocity profiles continue to deviate. Moreover, the crest temperature, $\Theta_k^+$ (inset of figure 4\emph{d}), increases in a log-linear manner, which for increasing $k^+$, is only consistent if $\Delta \Theta^+$ is approaching a constant value. These observations therefore all suggest that $\Delta \Theta^+$ remains constant in the fully rough regime.

\setlength{\unitlength}{1cm}
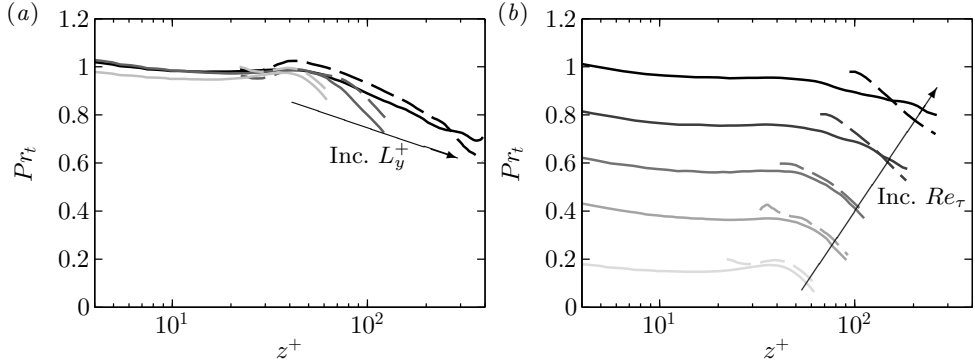
\begin{figure}
\centering
	\includegraphics[width=0.48\textwidth]{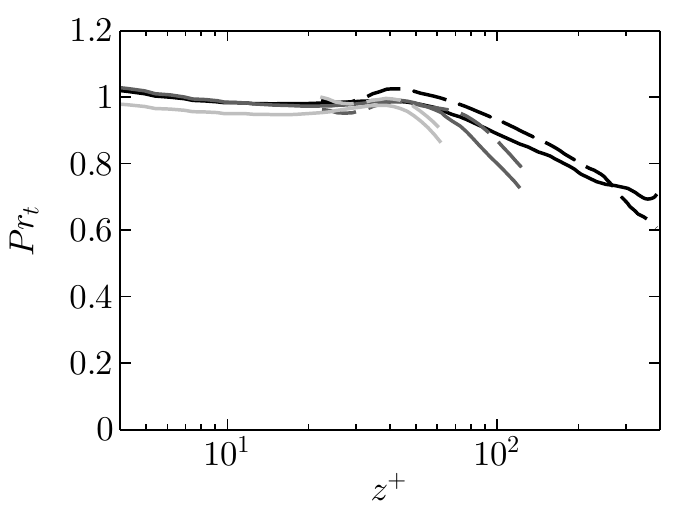}
	\put(-6.5,4.6){(\emph{a})}
	\put(-2.7,3.5){\vector(3,-1){2.2}}
	\put(-2.2,2.7){Inc.~$L_y^+$}
	\includegraphics[width=0.48\textwidth]{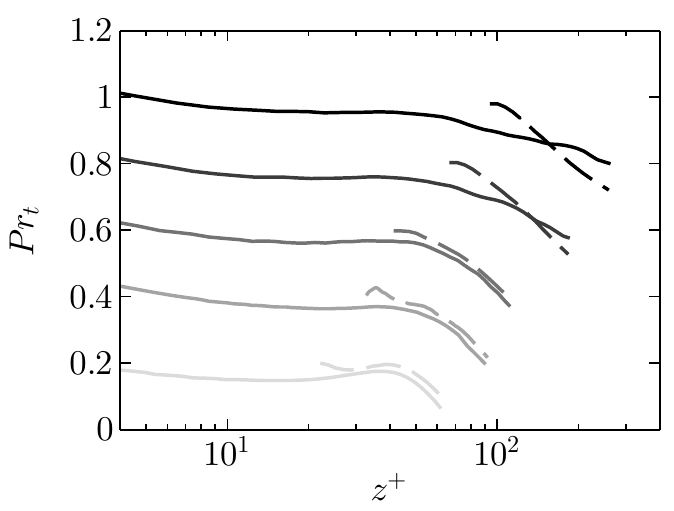}
	\put(-6.45,4.6){(\emph{b})}
	\put(-2.4,1.0){\vector(2,3){1.8}}
	\put(-1.4,2.15){Inc.~$Re_\tau$}
	\vspace{-0.5\baselineskip}
\caption{
Turbulent Prandtl number against wall-normal distance, $z^+$, for 
(\emph{a}) increasing channel width with $Re_\tau=395$ (set $A$, table \ref{tab:sims})
and
(\emph{b}) increasing Reynolds number (set $C$, table \ref{tab:sims}),
for smooth-wall (solid) and rough-wall (dashed) flows.
Data  are only shown from crest to critical height, $z_c^+$,
and in (\emph{b})  are staggered by $-0.2$ for decreasing $Re_\tau$.
}
	\label{fig:turbPr}
\end{figure}

An important quantity in heat transfer models is the turbulent Prandtl number, defined as the ratio of momentum and heat transfer eddy diffusivities,
\begin{equation}
Pr_t = \frac{\nu_t}{\alpha_t}=\frac{\langle\overline{u'w'}\rangle}{\langle\overline{w'\theta'}\rangle}\frac{\dd\Theta/\dd z}{\dd U/\dd z},
\end{equation}
where $\langle\overline{u'w'}\rangle$ is the Reynolds shear stress and $\langle\overline{w'\theta'}\rangle$ is the turbulent heat flux.
Under sufficiently high Reynolds and Peclet numbers, dimensional arguments predict that $Pr_t$ should be constant in the logarithmic layer \citep{Cebeci84}.
Indeed, \cite{Pirozzoli16} showed that the smooth-wall turbulent Prandtl number  is close to unity at the wall, before reducing to become approximately constant in the logarithmic layer, with $Pr_t\approx0.85$ over $z^+\gtrsim100$ and $z/h\lesssim0.5$. Studies of temporally developing boundary layers \citep{Kozul16} and statistically stationary homogeneous shear flows (which can be treated as a model for the logarithmic layer, see \citealt{Sekimoto16,Chung12})  also suggest $Pr_t\approx 1.0$.
In the wake region, $Pr_t$ reduces further towards values of 0.5, in line with free-shear flows, or the wake behind a bluff body \citep{Cebeci84}.

In figure \ref{fig:turbPr}(\emph{a}), the turbulent Prandtl number is shown for increasing channel widths at fixed $Re_\tau=395$ (set $A$, table \ref{tab:sims}), where the data are only shown for $z<z_c$. 
The full-span channel (black line) shows $Pr_t$ tending towards values of 0.85--0.9 within the logarithmic layer (around $z^+\approx 60$), although the
relatively low Reynolds numbers and short logarithmic layer do not enable the constant $Pr_t$ region to be obtained. Above the logarithmic layer, in the wake, magnitude of the full-span $Pr_t$ continue to decrease, as discussed above.
The narrowest channel with $L_y^+\approx 153$ (light grey solid line) has a turbulent Prandtl number that is slightly less than the wider span cases, even at the wall. If we consider the turbulent momentum and heat fluxes as integrals of their respective one-dimensional spanwise energy spectra,
$\langle\overline{u'w'}\rangle=\int_0^\infty E_{uw}\id k_y$ and $\langle\overline{w'\theta'}\rangle=\int_0^\infty E_{w\theta}\id k_y$, we see that some length scales (namely $k_y < 2\pi/L_y$) will be missing due to the use of the minimal channel, even below $z<z_c$. However, \cite{MacDonald17} showed that much of the dynamically relevant scales are still captured in these energy spectra, and we see the difference in $Pr_t$ between the different channel spans close to the wall in figure \ref{fig:turbPr}(\emph{a}) is fairly small. Moreover, $L_y^+$ increases with $Re_\tau$ so this effect is less significant for our larger Reynolds number cases.
 More noticeable is the reduction in $Pr_t$ at higher $z^+$ values (while still below $z_c^+$) for the minimal channels, relative to the full-span channel. This is likely due to the wake region, that now starts at $z_c^+$, encroaching into the immature logarithmic layer, leading to the early reduction in $Pr_t$. Note that $Pr_t$ is a highly sensitive measure of the relative slopes of the mean velocity and temperature profiles, and requires very high Reynolds and Peclet numbers to yield meaningful results. The differences in these mean profiles, $\Delta U^+$ and $\Delta \Theta^+$, meanwhile, are much less sensitive to the slopes and do not suffer as severely from this wake-encroachment issue. They are also the primary source of uncertainty in estimating the skin-friction and heat-transfer coefficients and, as shown above, are accurately measured with the minimal channel.

Figure \ref{fig:turbPr}(\emph{b}) shows the turbulent Prandtl number for increasing friction Reynolds numbers. We see that the rough-wall turbulent Prandtl number (dashed lines) is initially unity at the roughness crest (within the roughness sublayer) and therefore larger than that of the smooth wall at matched $z^+$.
For small $k^+$ values, the difference in the smooth- and rough-wall turbulent Prandtl numbers is minor and a reasonable collapse is observed.
For the two largest $k^+$ values which are nominally fully rough (dark grey dashed lines), the rough-wall $Pr_t$ at the crest is noticeably larger than the smooth-wall value at matched $z^+$.
It then rapidly reduces with wall-normal distance until it is slightly less than that of the smooth wall at $z_c^+$.
 From Townsend's outer-layer similarity hypothesis \citep{Townsend76}, we would expect the smooth- and rough-wall turbulent Prandtl numbers to eventually collapse in the logarithmic layer (with constant $Pr_t$), assuming sufficiently large $Re_\tau$ and wall-normal extent of the logarithmic layer.

\subsection{Skin-friction and heat-transfer coefficients}
\label{ssect:cfch}

\setlength{\unitlength}{1cm}
\begin{figure}
\centering
	\includegraphics[width=0.49\textwidth]{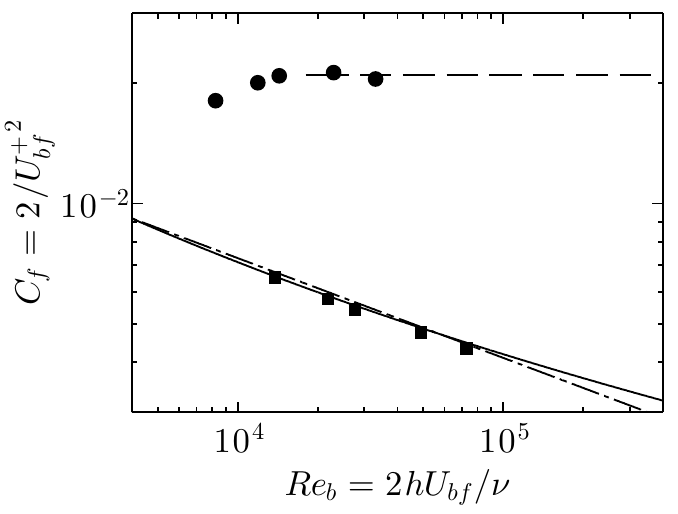}
	\includegraphics[width=0.49\textwidth]{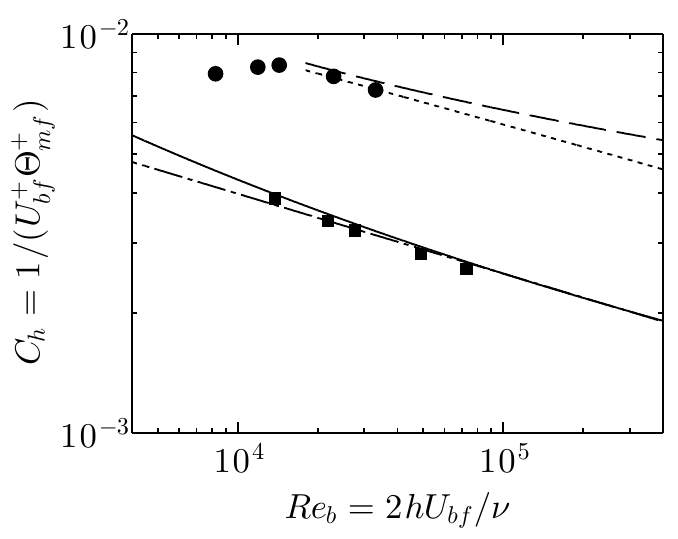}
	\put(-13.5,4.8){(\emph{a})}
	\put(-10.8,4.45){$k/h=1/18$}
	\put(-10.8,1.95){\rotatebox{-20}{Smooth}}
	\put(-9.90,1.15){D78}
	\put(-9.25,1.25){\vector(1,0){1.5}}
	\put(-6.6,4.8){(\emph{b})}
	\put(-3.3,4.65){\rotatebox{-12}{$k/h=1/18$}}
	\put(-3.7,2.8){\rotatebox{-18}{Smooth}}
	\put(-5.1,2.0){KCW05}
	\put(-2.3,3.85){\rotatebox{-15}{DS63}}
	\put(-4.8,2.3){\vector(0,1){1.2}}
	\\
	\includegraphics[width=0.49\textwidth]{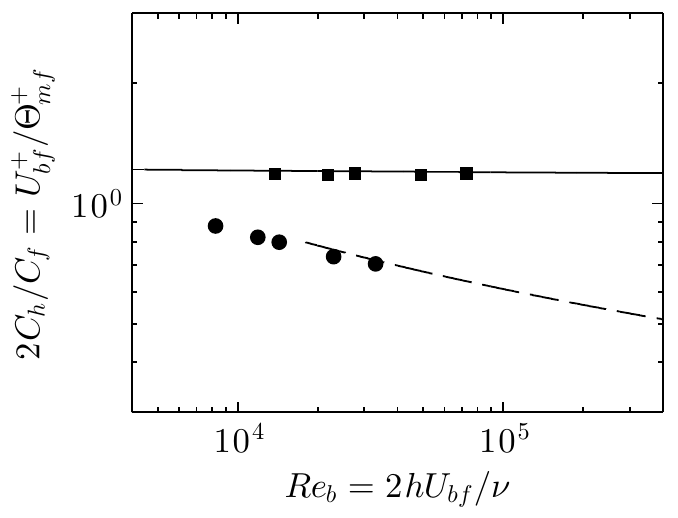}
	\includegraphics[width=0.49\textwidth]{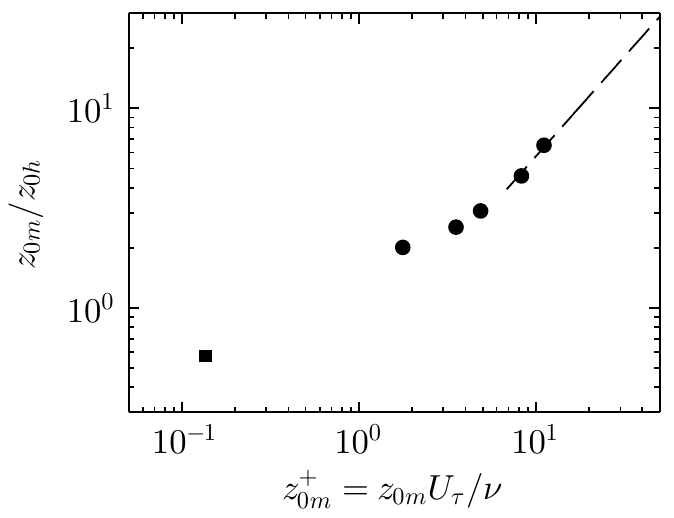}
	\put(-13.5,4.8){(\emph{c})}
	\put(-9,2.03){\rotatebox{-11}{$k/h=1/18$}}
	\put(-8.6,3.45){Smooth}
	\put(-6.0,4.8){(\emph{d})}
	\put(-1.75,3.50){\rotatebox{+48}{$k/h=1/18$}}
	\put(-4.4,1.48){Smooth}
	\vspace{-0.5\baselineskip}
\caption{
(\emph{a}) Skin-friction coefficient $C_f$, 
(\emph{b}) heat-transfer coefficient (Stanton number) $C_h$
and
(\emph{c}) ratio $2C_h/C_f$, as a function of bulk Reynolds number $Re_b=2U_{bf} h/\nu$.
(\emph{d}) Ratio of momentum and heat transfer roughness lengths $z_{0m}/z_{0h}$
against $z_{0m}^+$.
Symbols:
\protect\raisebox{-0.1ex}{\protect\scalebox{1.0}{$\color{black}\blacksquare$}}, smooth-wall data;
\protect\raisebox{-0.5ex}{\protect\scalebox{2.0}{$\color{black}{\bullet }$}}, rough-wall  data.
Theses are estimated by fitting full-span composite profiles to the minimal channel velocity and temperature profiles for $z>z_c$.
Line styles:
\dashDotLine{black}, smooth-wall power-law correlations \citep{Dean78,Kays05};
\solidLine{black}, smooth-wall  log-law estimate using (\ref{eqn:UbLambert}) and (\ref{eqn:Thetab_Ub});
\dotLine{black}, fully rough heat transfer model (\ref{eqn:CHdipprey}) of \cite{Dipprey63};
\dashLine{black}, fully rough log-law estimate (\ref{eqn:Stanton_FR}--\ref{eqn:z0h}).
}
	\label{fig:CfCh}
\end{figure}

The skin-friction and heat-transfer coefficients, $C_f=2/{U_{bf}^{+2}}$ and $C_h=1/(U_{bf}^+ \Theta_{mf}^+)$, are given in figure \ref{fig:CfCh}(\emph{a}) and (\emph{b}) for the present smooth-wall and rough-wall data with $k/h=1/18$ (symbols). Here,  the expected full-span velocity ($U_f$) and temperature ($\Theta_f$) profiles are computed by fitting the composite profile of \cite{Nagib08} for full-span channel flow to the minimal channel data for $z>z_c$. We use slope coefficients of $\kappa_m=0.4$ and $\kappa_h=0.46$ \citep{Pirozzoli16} and the same empirical wake function, $\Psi$, of \cite{Nagib08} for both velocity and temperature profiles. However, the wake parameter, $\Pi$, is set to 0.08 for velocity and 0.03 for temperature, where these constants comes from fitting the outer-layer composite profile to the data of \cite{Bernardini14} and \cite{Pirozzoli16} for velocity and temperature, respectively. The offsets, $A_m$ and $A_h$, are set such that the profile is continuous at $z_c$.

Empirical correlations are also given in figure \ref{fig:CfCh}(\emph{a},\emph{b}). The power law of \cite{Dean78} for the smooth-wall skin-friction coefficient,
$C_{fs} = 0.073Re_b^{-1/4}$, is provided (dash-dotted line in figure \ref{fig:CfCh}\emph{a}), as well as the Prandtl--von K{\'a}rm{\'a}n logarithmic skin-friction law (solid line). The latter comes from integrating the logarithmic smooth-wall mean velocity profile (as in (\ref{eqn:logU}) with $\Delta U^+=0$) across the entire channel to obtain an implicit equation,
\begin{equation}
\label{eqn:UbImplicit}
U_{bs}^+=\frac{1}{\kappa_m}\log\left(\frac{\frac{1}{2}Re_b}{U_{bs}^+}\right)-\frac{1}{\kappa_m}+A_m.
\end{equation}
This can be solved for $C_{fs}$ in terms of $Re_b$ using the product logarithm (or Lambert's $\mathcal{W}$-function), resulting in
\begin{equation}
\label{eqn:UbLambert}
U_{bs}^+=\sqrt{\frac{2}{C_{fs}}}=\frac{1}{\kappa_m}\mathcal{W}\left(\frac{1}{2}Re_b\kappa_m e^{(A_m\kappa_m-1)}\hspace{0.05cm}\right).
\end{equation}
While both the power-series and log-law skin-friction coefficient models agree well with the present smooth-wall data at moderate Reynolds number (figure \ref{fig:CfCh}\emph{a}), recent studies of smooth-wall channel flow have shown that the log-law equation agrees better with high Reynolds number data than the smooth-wall power-series correlations \citep{Schultz13,Bernardini14}.

Similarly, the smooth-wall power-series heat-transfer coefficient of \cite{Kays05} is given in figure \ref{fig:CfCh}(\emph{b}), $C_{hs}=0.021Re_b^{-0.2}Pr^{-0.5}$. As with the velocity profile, the smooth-wall logarithmic temperature profile (as in (\ref{eqn:logT}) with $\Delta \Theta^+=0$) can be integrated across the entire channel to obtain a similar expression to (\ref{eqn:UbImplicit}), with
\begin{equation}
\label{eqn:Thetab_Ub}
\Theta_{as}^+=\frac{1}{\kappa_h}\log\left(\frac{\frac{1}{2}Re_b}{U_{bs}^+}\right)-\frac{1}{\kappa_h}+A_h,
\end{equation}
where $\kappa_h\approx0.46$ and $A_h\approx3.2$ for the present $Pr=0.7$ flow (figure \ref{fig:veltempIncK}\emph{b}). 
Hence, the smooth-wall Stanton number $C_{hs}=1/(U_{bs}^+ \Theta_{ms}^+)$ can be estimated using (\ref{eqn:UbLambert}) and  (\ref{eqn:Thetab_Ub}), 
where (\ref{eqn:Tma}) is used to get the mixed-mean temperature, $\Theta_{m}^+$, in terms of the arithmetic mean temperature, $\Theta_{a}^+$.
 At moderate Reynolds numbers, both the power-series correlation of \cite{Kays05} and the log-law formulas show good agreement with the present data for the smooth-wall heat-transfer coefficient (squares in figure \ref{fig:CfCh}\emph{b}). At higher Reynolds numbers however, we might expect the log-law equations to perform better like they do with the skin-friction coefficient.

The present rough-wall skin-friction coefficient (circles in figure \ref{fig:CfCh}\emph{a}) is seen to initially increase in value in the transitionally rough regime. However, for sufficiently large bulk Reynolds number it is tending towards a constant value of $C_{f_{FR}}\approx0.021$ (dashed line). This indicates that the flow is approaching the asymptotic fully rough state. The rough-wall heat-transfer coefficient (circles in figure \ref{fig:CfCh}\emph{b}), meanwhile, increases to a maximum in the transitionally rough regime, before monotonically reducing in the fully rough regime. 
The dotted line here shows the heat transfer model of \citet[][eq. 28]{Dipprey63}, in which the fully rough (subscript $FR$) heat-transfer coefficient for a pipe was given as
\begin{equation}
\label{eqn:CHdipprey}
C_{h_{FR}}=\frac{\frac{C_{f_{FR}}}{2}}{1+\sqrt{\frac{C_{f_{FR}}}{2}}\left[k_f \left(Re_b\sqrt{\frac{C_{f_{FR}}}{2}}\frac{k_s}{2h}\right)^{0.2}Pr^{0.44}-8.48\right]},
\end{equation}
where $k_s/(2h)=\exp\left[\kappa_m(3.0-\sqrt{2/C_{f_{FR}}})\right]$ is the blockage in terms of the equivalent sand-grain roughness.
The constant $k_f$ is roughness dependent, where \cite{Dipprey63} suggested $k_f=5.19$ for granular roughness. Here, we use $k_f=5.6$, obtained from a least squares fit to the fully rough data ($k^+\gtrsim 66$). While (\ref{eqn:CHdipprey}) was developed for pipe flow, we see it correctly predicts the trend of $C_h$ reducing with Reynolds number for the present channel flow cases.

Alternatively, we can use the assumed logarithmic velocity and temperature profiles to obtain an expression for the fully rough heat transfer coefficient (Stanton number). Starting from the definition of the Stanton number and (\ref{eqn:Tma}) we get,
\begin{equation}
\label{eqn:Stanton_FR}
C_{h_{FR}}=\frac{1}{U_{b_{FR}}^+\Theta_{m_{FR}}^+}=\frac{\kappa_m\kappa_h}{1+\kappa_m\kappa_h U_{b_{FR}}^+\Theta_{a_{FR}}^+},
\end{equation}
where $U_{b_{FR}}^+$ is the fully rough bulk velocity, which can be obtained by integrating the logarithmic velocity profile (\ref{eqn:logU}) with $\Delta U^+=\kappa_m^{-1}\log(k^+)+C$, to yield a constant,
\begin{eqnarray}
U_{b_{FR}}^+ &=& A_m-C-\frac{1}{\kappa_m}\left(1+\log\left(\frac{k}{h}\right)\right)\label{eqn:UbFR_C}\\
                          &=& C_N-\frac{1}{\kappa_m}\left(1+\log\left(\frac{k_s}{h}\right)\right),
\label{eqn:UbFR_ks}
\end{eqnarray}
 and $\Theta_{a_{FR}}^+$ can be obtained by integrating the logarithmic rough-wall temperature profile, as in (\ref{eqn:logT}) with constant temperature difference $\Delta \Theta_{FR}^+$, to yield
\begin{eqnarray}
\Theta_{a_{FR}}^+ &=& \frac{1}{\kappa_h}\log\left(\frac{\frac{1}{2}Re_b}{U_{b_{FR}}^+}\right)-\frac{1}{\kappa_h}+A_h-\Delta \Theta_{FR}^+ .
\label{eqn:dtheta_ks}
\end{eqnarray}
Here, the log-law constants, $A_m\approx 5.0$, $\kappa_m \approx 0.4$, $\kappa_h\approx0.46$ and $A_h(Pr=0.7)\approx3.2$ are all known and $C_N=8.5$ is Nikuradse's constant. This means that we only need the roughness function offset, $C$, from $\Delta U^+=\kappa_m^{-1}\log(k^+)+C$ in (\ref{eqn:UbFR_C}), or alternatively the equivalent sand-grain roughness, $k_s$, in the form given by (\ref{eqn:UbFR_ks}), as well as the constant $\Delta \Theta_{FR}^+$ to predict the Stanton number at any Reynolds number in the fully rough regime. These dynamical parameters  ($k_s/k$ and $\Delta \Theta_{FR}$) are geometry dependent and must be measured in a dynamic procedure. Here, the  unknowns can be determined using the minimal channel technique; for the present sinusoidal roughness, these values are $k_s/k\approx4.1$ and $\Delta {\Theta_{FR}}^+\approx 4.4$ (figure \ref{fig:DUDT}).

The empirical heat-transfer model of \cite{Dipprey63} in (\ref{eqn:CHdipprey}) as well as the integrated log-law equations in (\ref{eqn:UbFR_ks}--\ref{eqn:dtheta_ks}) are shown in figure \ref{fig:CfCh}(\emph{b}), along with the present data (circles). Both models show good agreement with the present data at moderate Reynolds number, with the heat-transfer coefficient reducing with Reynolds number. However, the advantage of the log-law equations is that the only modelling assumption made is that the temperature profile follows a logarithmic profile across the entire channel. This is phenomenologically consistent with our understanding of forced convection wall turbulence and does not require any \emph{ad hoc} terms like $k_f$ in (\ref{eqn:CHdipprey}). The integrated log-law equations can also be extended to include the effect of the wake through additive constants $L_m=\int_0^h2\Pi_m/(\kappa_m)\Psi\id z$ and $L_h=\int_0^h2\Pi_h/(\kappa_h)\Psi\id z$. These are small for channel flows so do not significantly alter the results here, but are larger for pipes and boundary layers.

Figure \ref{fig:CfCh}(\emph{c}) shows the ratio of the heat-transfer and skin-friction coefficients, $2C_h/C_f$, for the present data as well as the log-law equations.
We see that the smooth-wall ratio is constant with Reynolds number, in support of the Reynolds analogy in which momentum transfer is proportional to heat transfer. In contrast, the rough-wall ratio reduces with Reynolds number, indicating that the Reynolds analogy for rough-wall flow is breaking down for these bulk measures of momentum and heat transfer. Note that this is only in regard to the bulk quantities; the analogy may still hold within the flow when looking at quantities such as the turbulent diffusivity for momentum and heat at a given wall-normal location.
 In an engineering sense, this ratio of coefficients can be regarded as the heat-transfer rate per unit pumping power \citep{Dipprey63}, or that the roughness is advantageous for heat transfer applications when $(C_{hr}/C_{fr})>(C_{hs}/C_{fs})$. 
Evidently, the present roughness and flow conditions are not as efficient as the smooth-wall flow. However, \cite{Dipprey63} noted that increasing the Prandtl number above $Pr\gtrsim3$ may enable advantageous heat transfer for the rough-wall flow to be obtained. This favourable condition typically only occurs when the flow is in the transitionally rough regime and the rough-wall heat-transfer coefficient has reached a maximum.
Note that there are also alternative measures for the performance of heat transfer systems depending on the engineering design constraints present \citep{Webb81}.

Finally, in meteorology, the logarithmic velocity profile (\ref{eqn:logU}) is often given as $U^+=(1/\kappa_m) \log(z/z_{0m})$, where $z_{0m}$ is the roughness length for momentum transfer. This is related to the equivalent sand-grain roughness by a constant, $k_s=\exp[\kappa_mC_N]z_{0m}\approx 30 z_{0m}$ \citep{Jimenez04}. In the fully rough regime, $k_s/k$ and $z_{0m}/k$ are constants that only depend on the roughness geometry. In a similar manner, the logarithmic temperature profile is defined as $\Theta^+=(1/\kappa_h)\log(z/z_{0h})$ where $z_{0h}$ is the roughness length for heat transfer \citep[e.g.][]{Owen63,Chamberlain66,Wood91}. Relating this to (\ref{eqn:logT}) in the fully rough regime, we obtain
\begin{equation}
\label{eqn:z0h}
z_{0h}^+=\exp[-\kappa_h(A_h-\Delta \Theta_{FR}^+)],
\end{equation}
where for the present roughness and flow conditions, $z_{0h}^+\approx 1.7$. Importantly, we see that this inner-normalised roughness length for heat transfer (or any passive scalar) is constant and does not depend on the roughness height, $k$, like its momentum counterpart.
This is the same form as suggested by \cite{Garratt73}, although there the authors used a  slightly different roughness-independent constant, $z_{0h}^+\approx1/(\kappa Pr) \approx 3.5$, where 
the heat transfer slope constant was assumed to be equal to the momentum constant with $\kappa=0.41$. In any case, this constant value is larger than the smooth-wall constant, which would be $z_{0hs}^+=\exp(-\kappa_hA_h)\approx0.23$.
\cite{Garratt73} presented data from a collection of experimental studies which for moderate Reynolds numbers provided strong support for the rough-wall constant $z_{0h}^+$ form above.
 However,  at higher Reynolds numbers of $Re_\tau\gtrsim 10^4$ some of the experimental data for the ratio $z_{0m}/z_{0h}$ increased faster than was predicted by using a constant $z_{0h}^+$ (i.e. $z_{0m}/z_{0h}\approx\kappa Pr z_{0m}^+$). In the present formulation using (\ref{eqn:z0h}), this ratio would take the form  $z_{0m}/z_{0h}\approx \exp[\kappa_h(A_h-\Delta \Theta_{FR}^+)]z_{0m}^+=Bz_{0m}^+$ where the constant $B\approx0.58$ for the present sinusoidal roughness. This is shown in figure \ref{fig:CfCh}(\emph{d}) and agrees well with the data for the present moderate Reynolds numbers.
  A more rapid increase in the ratio   for very large Reynolds numbers would correspond to $\Delta \Theta_{FR}^+$ decreasing.
More complex models  for $z_{0m}/z_{0h}$ involving non-linear expressions of $z_{0m}^+$ have been suggested to account for this increase \citep[e.g.][]{Owen63,Brutsaert75,Andreas87}. 
However, there is large scatter in the experimental data for high Reynolds number flows, with some data even suggesting $z_{0m}/z_{0h}$ eventually reduces \citep{Garratt73}, making it difficult to assess the true behaviour of $z_{0m}/z_{0h}$ (or $\Delta \Theta_{FR}^+$). For completeness,  while an equivalent sand-grain roughness for heat transfer, $k_{sh}$, does not appear to be used in the literature, it would take the form $k_{sh}=\exp[\kappa_h C_{Nh}]z_{0h}$, where $C_{Nh}$ would be the heat-transfer constant for Nikuradse's sand grain roughness which is currently unknown.

\subsection{Wall fluxes}
\label{ssect:walldrag}

\setlength{\unitlength}{1cm}
\begin{figure}
\centering
	\includegraphics[width=0.47\textwidth]{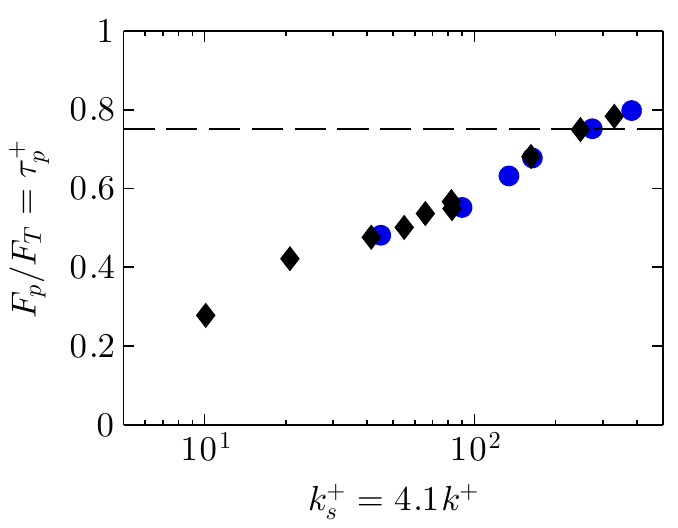}
	\includegraphics[width=0.482\textwidth,trim=-8 0 0 0,clip=true]{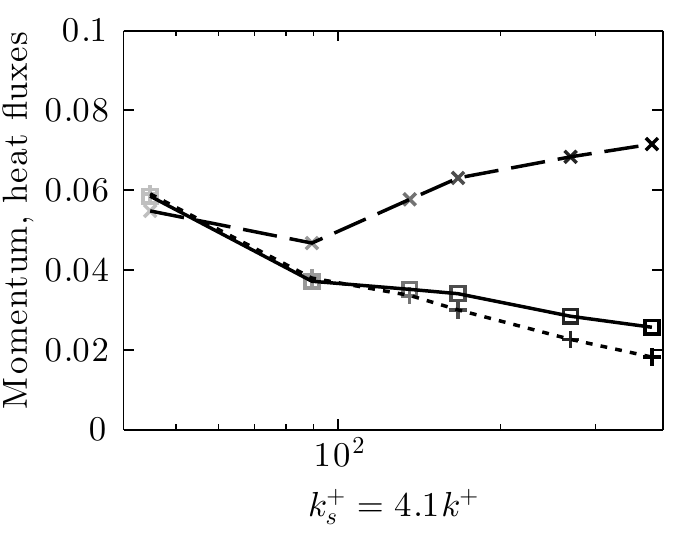}
	\put(-11.55,3.85){Fully rough}
	\put(-11.55,3.4){Transitionally rough}
	\put(-12.9,4.6){(\emph{a})}
	\put(-6.65,4.6){(\emph{b})}
	\put(-1.75,1.65){\rotatebox{-13}{Viscous}}
	\put(-1.5,2.25){\rotatebox{-12}{Heat}}
	\put(-1.75,3.45){\rotatebox{+11}{Pressure}}
	\vspace{-0.5\baselineskip}
\caption{(Colour online) (\emph{a}) Ratio of pressure to total drag force, $F_p/F_T$, against equivalent sand-grain roughness, $k_s^+=4.1k^+$. 
Symbols:
\protect\raisebox{-0.5ex}{\protect\scalebox{2.0}{$\color{myblue}{\bullet }$}}, present data;
$\color{black}\blacklozenge$, pipe flow with the same sinusoidal roughness geometry \citep{Chan15}.
(\emph{b}) Momentum and heat fluxes non-dimensionalised on crest velocity and temperature, $U_k$ and $\Theta_k$.
Line styles:
\dashLine{black}, pressure drag contribution $\tau_p/((1/2)\rho U_k^2)$;
\dotLine{black}, viscous drag contribution $\tau_\nu/((1/2)\rho U_k^2)$;
\solidLine{black}, wall heat flux $q_w/(\rho c_p U_k \Theta_k)$.
}
	\label{fig:FpFT}
\end{figure}
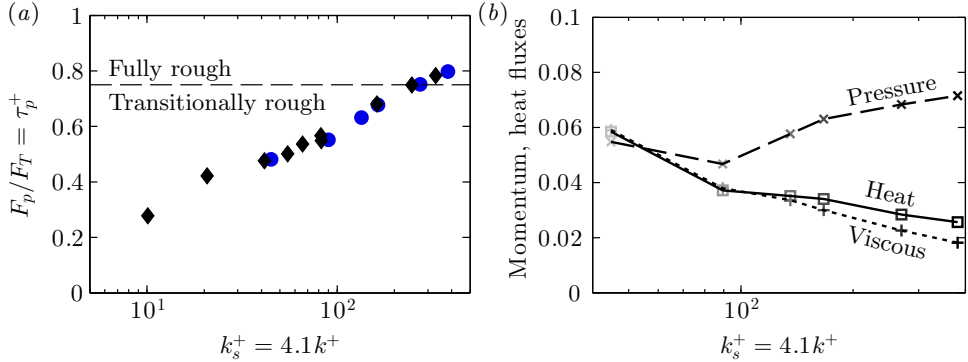

The ratio of pressure drag to total drag (pressure plus viscous drag) is shown in figure \ref{fig:FpFT}(\emph{a}). Equivalently, the total stress can be decomposed into pressure and viscous stress contributions, $\tau_w = \tau_p+\tau_v$, so that this figure also represents the inner-normalised pressure stress, $\tau_p^+$. Also shown in this figure is the pipe-flow data of \cite{Chan15} for the same sinusoidal roughness geometry, with good agreement observed between the two data sets. From the roughness function values shown in figure \ref{fig:DUDT}, it appears that the two roughness cases where $k_s^+\gtrsim275$ are tending towards the fully rough asymptote of $\kappa_m^{-1}\log(k_s^+)+A_m-8.5$, where $\Delta \Theta^+$ is approximately constant. In figure \ref{fig:FpFT}, this then corresponds to the pressure drag being larger than approximately 75\% of the total drag force in the fully rough regime. 
Comparing figures \ref{fig:DUDT} and \ref{fig:FpFT}(\emph{a}) shows that while the temperature difference $\Delta \Theta^+$ has become approximately constant for $k_s^+\gtrsim 275$,
the pressure drag continues to increase in a log-linear manner with $k_s^+$.
There is no distinct change in this trend between the transitionally and fully rough regimes and the viscous drag is still somewhat significant at 25\% of the total drag force.


Figure \ref{fig:FpFT}(\emph{b}) shows the momentum and heat fluxes non-dimensionalised with the crest velocity and temperature, $U_k$ and $\Theta_k$. The momentum flux has been decomposed into its pressure and viscous drag contributions (dashed and dotted lines, respectively), where the sum of these two contributions gives the drag coefficient, $C_{dk}=2/U_k^{+2}\approx0.09$, which was seen to be approximately constant at in the inset of figure \ref{fig:veltempIncK}(\emph{c}). 
Note that the reduction in the pressure drag component for $k_s^+\approx 90$ is due to the slight increase in $U_k^+$ for this $k^+$; the ratio of pressure to total drag force (figure \ref{fig:FpFT}\emph{a}) increases with $k_s^+$ as expected.
Alongside the momentum fluxes shown in figure \ref{fig:FpFT}(\emph{b}) is the wall heat flux, $q_w/(\rho c_p U_k \Theta_k)=1/(U_k^+\Theta_k^+)$, which follows a similar trend to the viscous  contribution. This suggests that in rough-wall forced convection, the Reynolds analogy can only be used to relate the heat flux to the viscous drag contribution of the momentum flux. The analogy does not hold for the overall momentum flux due to the pressure-drag contribution, and explains why the ratio $2C_h/C_f$ (figure \ref{fig:CfCh}\emph{c}) reduces with Reynolds number for the rough-wall case.

\setlength{\unitlength}{1cm}
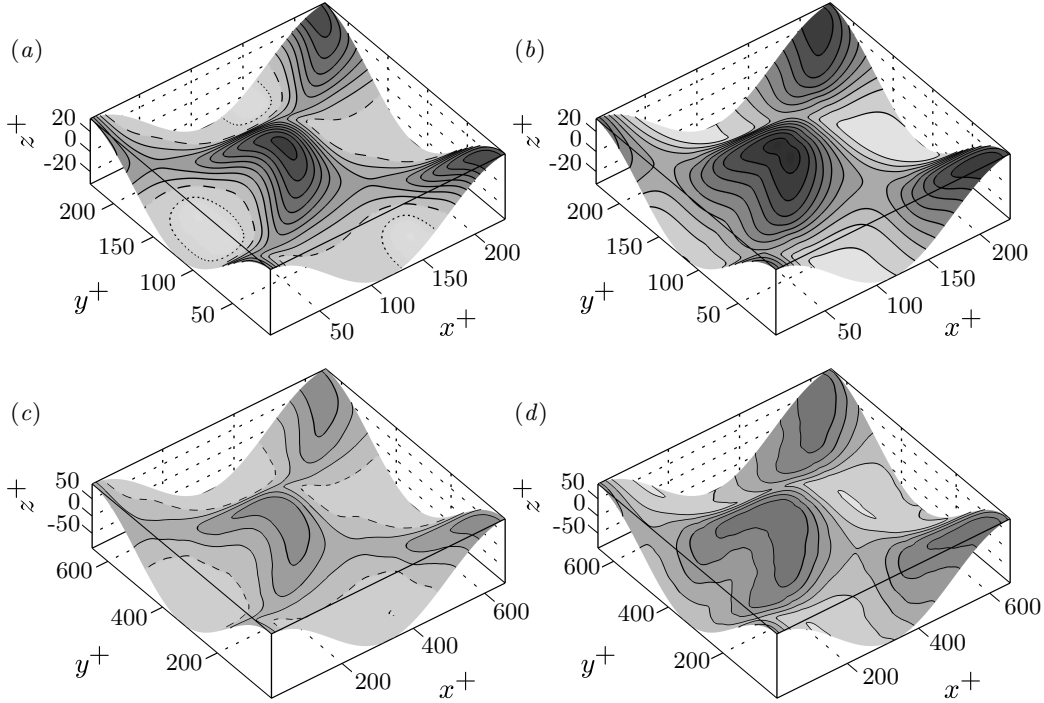
\begin{figure}
\centering
	\includegraphics[width=0.49\textwidth]{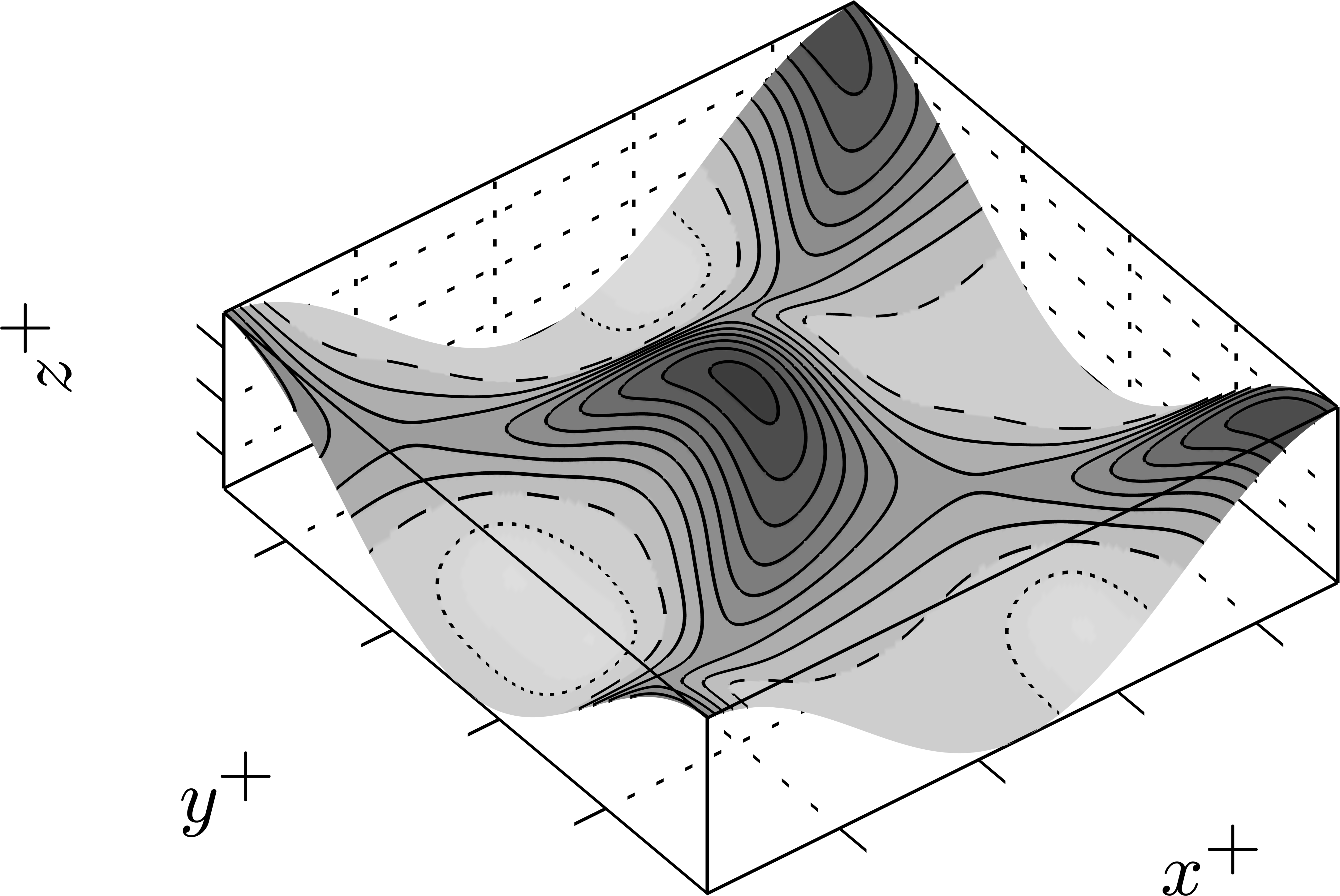}
	\includegraphics[width=0.49\textwidth]{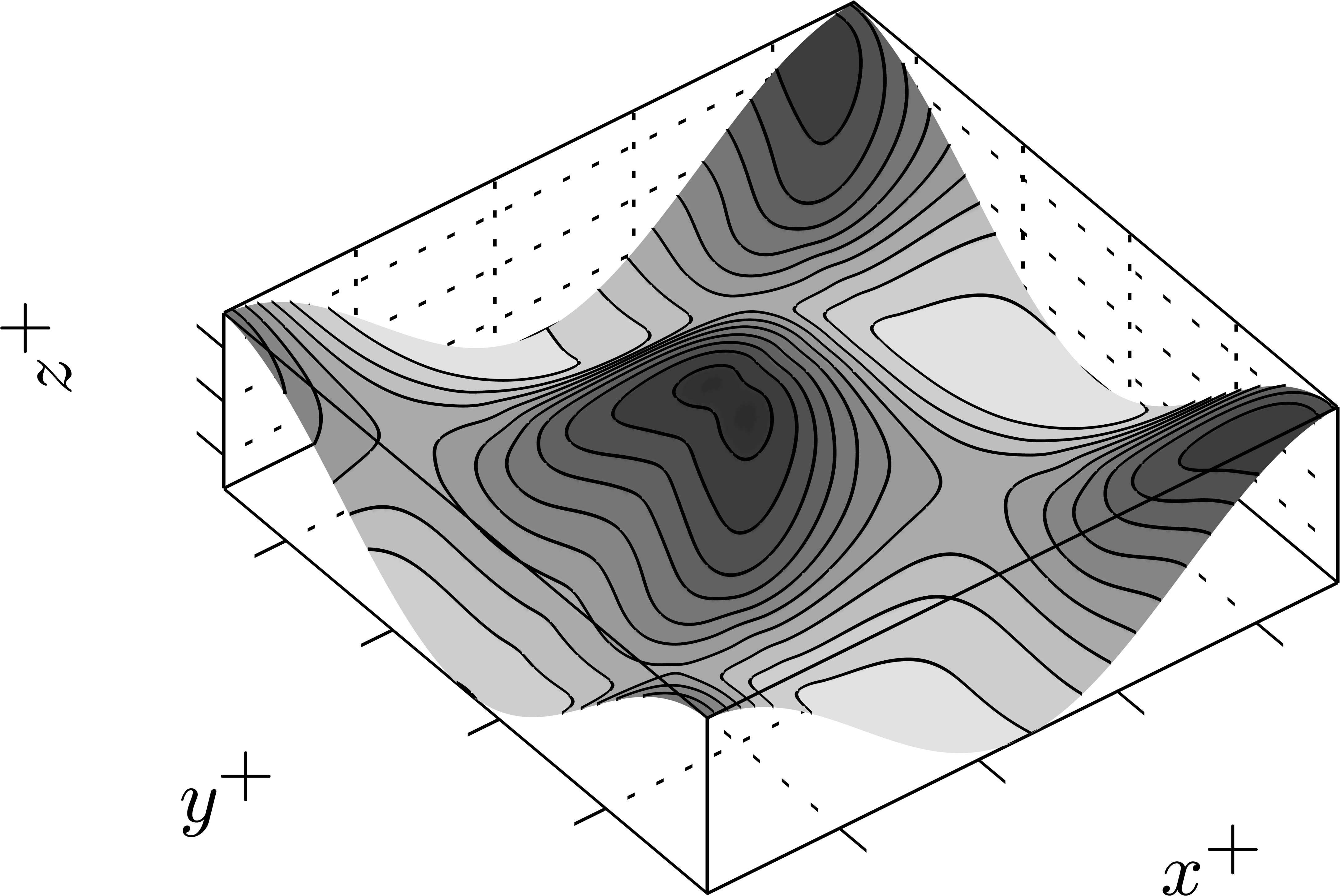}
	\put(-13.3,3.7){(\emph{a})}
	\put(-7.0,1.0){200}
	\put(-7.7,0.65){150}
	\put(-8.4,0.3){100}
	\put(-9.05,-0.05){50}
	\put(-12.65,1.55){200}
	\put(-12.13,1.08){150}
	\put(-11.58,0.6){100}
	\put(-10.9,0.18){50}	
	\put(-12.85,2.2){-20}
	\put(-12.58,2.52){0}
	\put(-12.75,2.8){20}
	\put(-6.6,3.7){(\emph{b})}
	\put(-0.25,1.0){200}
	\put(-0.95,0.65){150}
	\put(-1.65,0.3){100}
	\put(-2.3,-0.05){50}
	\put(-5.90,1.55){200}
	\put(-5.38,1.08){150}
	\put(-4.83,0.6){100}
	\put(-4.18,0.18){50}	
	\put(-6.13,2.2){-20}
	\put(-5.86,2.52){0}
	\put(-6.03,2.8){20}
\\
\vspace{+1.0\baselineskip}
	\includegraphics[width=0.49\textwidth]{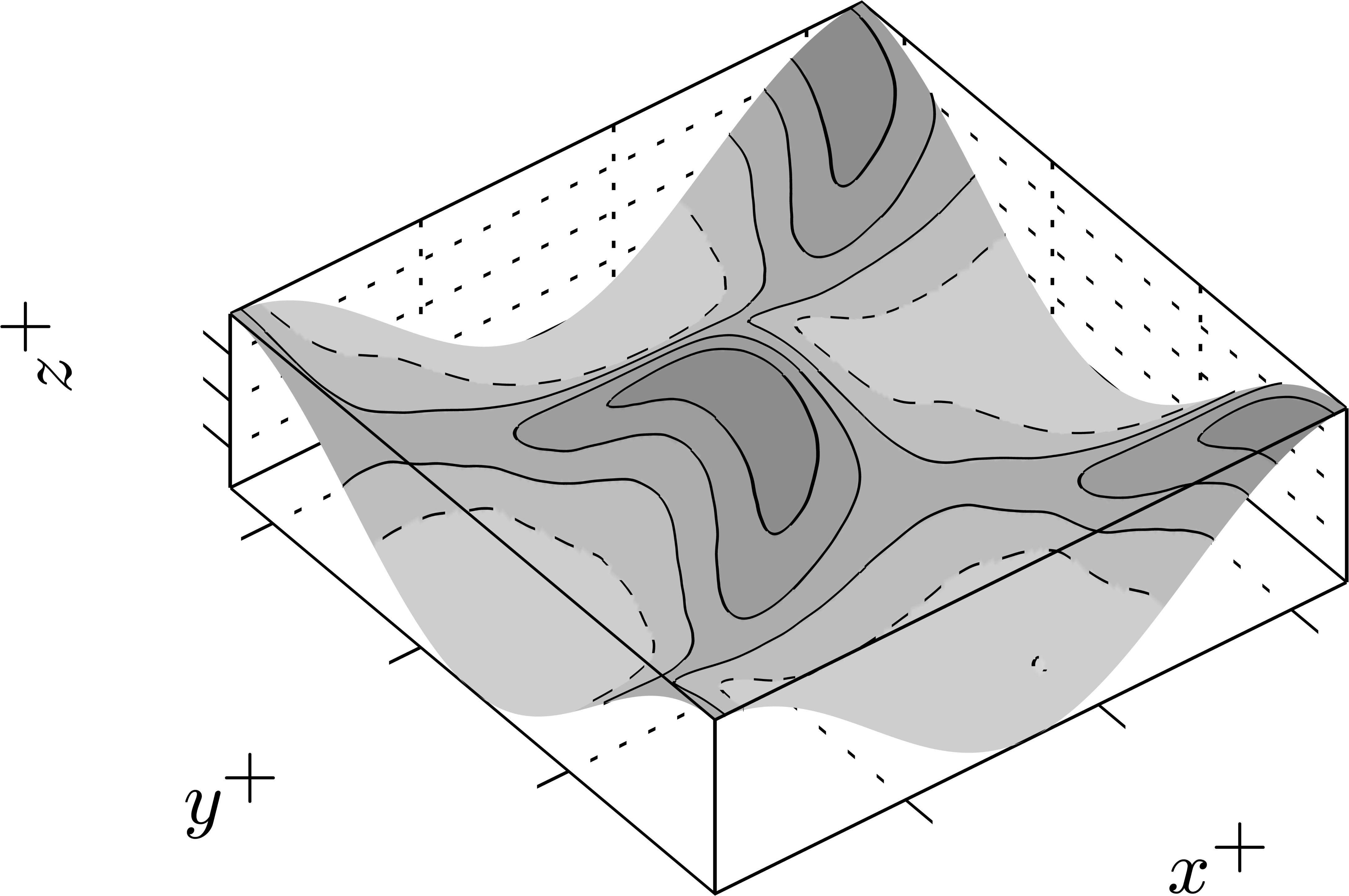}
	\includegraphics[width=0.49\textwidth]{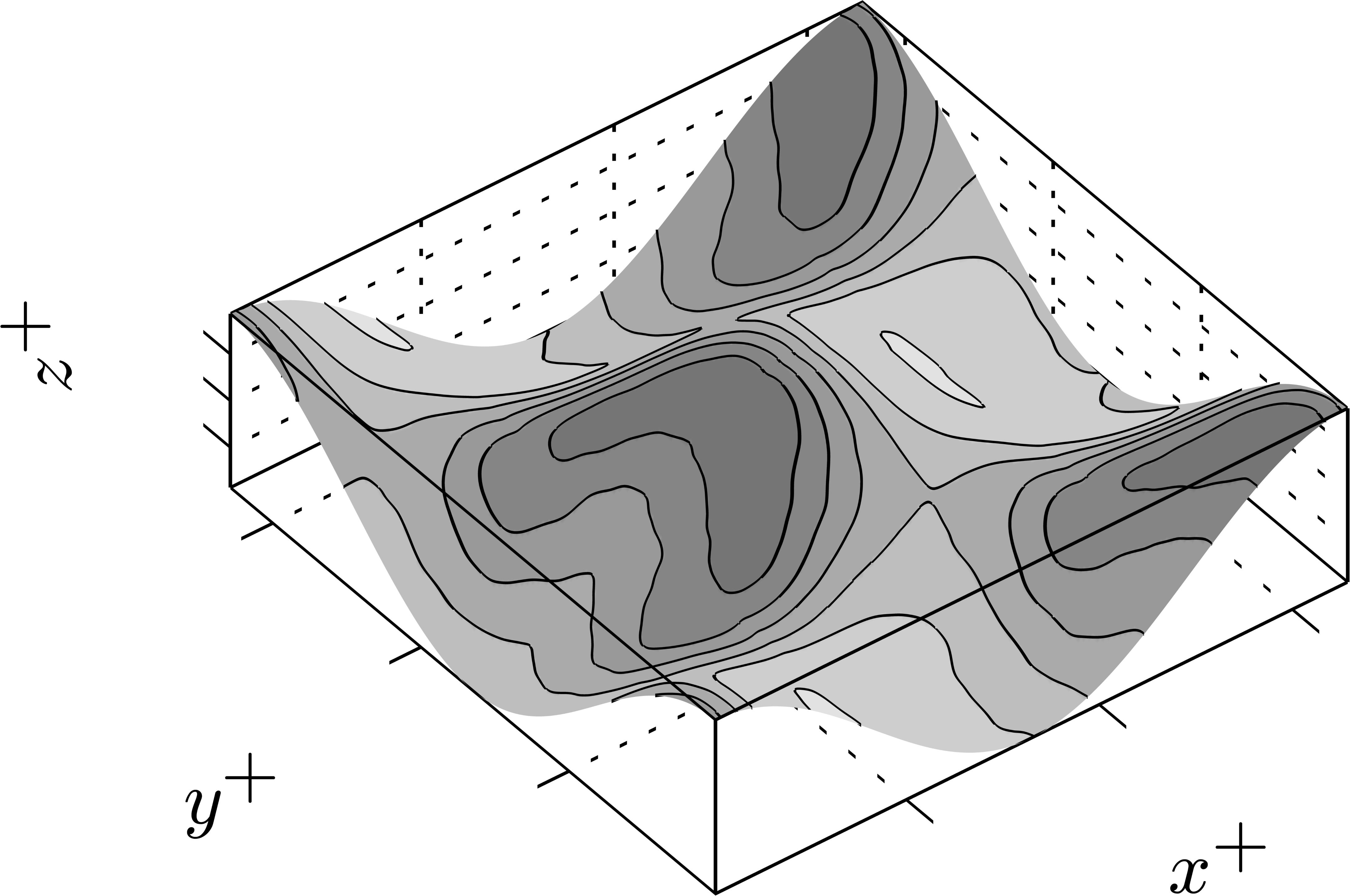}
	\put(-13.3,3.7){(\emph{c})}
	\put(-6.88,1.05){600}
	\put(-7.83,0.60){400}
	\put(-8.72,0.12){200}
	\put(-12.69,1.6){600}
	\put(-11.96,0.98){400}
	\put(-11.23,0.38){200}	
	\put(-12.8,2.22){-50}
	\put(-12.53,2.52){0}
	\put(-12.7,2.8){50}
	\put(-6.6,3.7){(\emph{d})}
	\put(-0.08,1.12){600}
	\put(-1.08,0.64){400}
	\put(-1.97,0.19){200}
	\put(-5.94,1.65){600}
	\put(-5.22,1.03){400}
	\put(-4.48,0.4){200}	
	\put(-6.08,2.2){-50}
	\put(-5.81,2.5){0}
	\put(-5.98,2.78){50}
	\vspace{-0.0\baselineskip}
\caption{(\emph{a}, \emph{c}) Viscous stress and (\emph{b}, \emph{d}) wall heat flux for roughness with 
(\emph{a}, \emph{b}) $k^+\approx 33$ (transitionally rough) and (\emph{c}, \emph{d}) $k^+\approx 93$ (fully rough). 
Averaged over time and for each repeating roughness element.
Dashed contour shows zero stress (recirculation) region, dotted contour in (\emph{a}) shows negative stress with value $-0.2U_\tau^2$. Contours are equally spaced with intervals of (\emph{a}, \emph{c}) $0.2U_\tau^2$ and (\emph{b},\emph{d}) $0.2\Theta_\tau U_\tau$.
Flow is from lower left to upper right. 
}
	\label{fig:tauPhiContour}
\end{figure}

The average viscous stress and heat flux on the rough wall are shown in figure \ref{fig:tauPhiContour} for transitionally rough flow ($k^+\approx 33$, \emph{a},\emph{b}) and nominally fully rough flow ($k^+\approx 93$, \emph{c},\emph{d}). These have been temporally averaged
from data outputted approximately every $35\nu/U_\tau^2$ as well as averaged over each repeating roughness element in the domain.
 The viscous stress (\emph{a},\emph{c}) is strongest at the roughness crests as there is high-speed fluid passing over the crests, resulting in substantial shear. This region of strong viscous stress extends down towards the saddles between neighbouring crests. A region of negative stress forms on the lee side of the roughness due to recirculation behind the roughness crests. 
The negative stress is stronger for the transitionally rough flow however the area covered by this negative stress (the dashed contour line) is approximately constant, at 40\% of the roughness plan area for both the transitionally (figure \ref{fig:tauPhiContour}\emph{a}) and nominally fully rough (\emph{c}) flows.
The heat flux (figure \ref{fig:tauPhiContour}\emph{b},\emph{d}) appears similar to the viscous stress contours, with strong heat flux localised to the roughness crests and along the saddles.   This occurs because these regions are exposed to faster flow and higher fluid temperature, leading to enhanced viscous stress and heat flux.

\setlength{\unitlength}{1cm}
\begin{figure}
\centering
	\includegraphics[trim=3 9 5 0,clip = true,scale = 0.88]{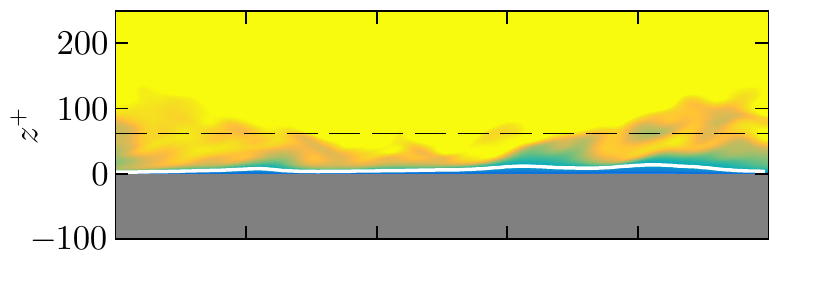}
	\put(-7.1,2.1){(\emph{a})}
	\includegraphics[trim=22 9 0 0,clip = true,scale = 0.88]{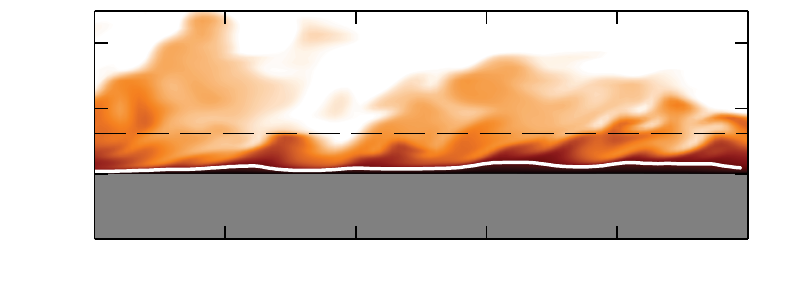}
	\put(-4.27,1.7){\colorbox{white}{Smooth wall, $Re_\tau\approx395$}}
	\\
	\includegraphics[trim=3 9 5 0,clip = true,scale = 0.88]{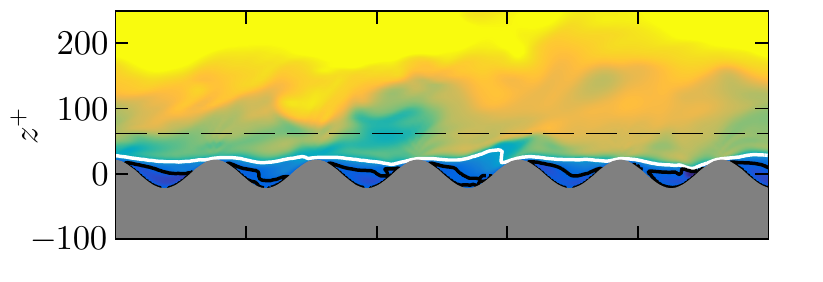}
	\put(-7.1,2.1){(\emph{b})}
	\includegraphics[trim=22 9 0 0,clip = true,scale = 0.88]{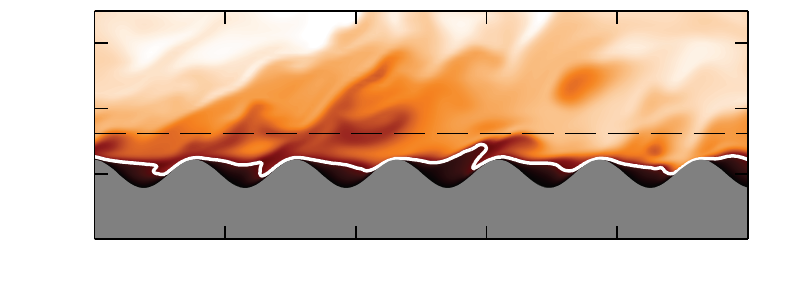}
	\put(-3.65,1.7){\colorbox{white}{$k^+\approx21$, $Re_\tau\approx395$}}
	\\
	\includegraphics[trim=3 9 5 0,clip = true,scale = 0.88]{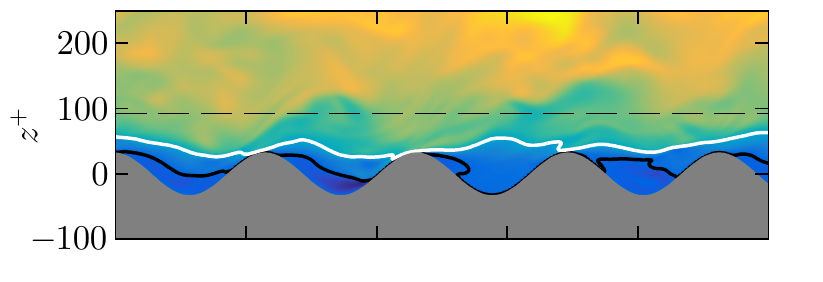}
	\put(-7.1,2.1){(\emph{c})}
	\includegraphics[trim=22 9 0 0,clip = true,scale = 0.88]{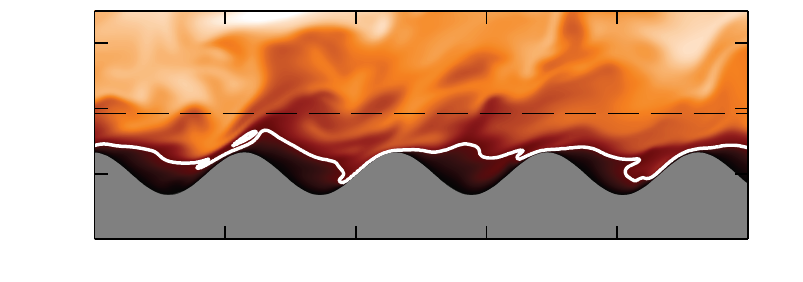}
	\put(-3.65,1.7){\colorbox{white}{$k^+\approx33$, $Re_\tau\approx590$}}
	\\
	\includegraphics[trim=3 9 5 0,clip = true,scale = 0.88]{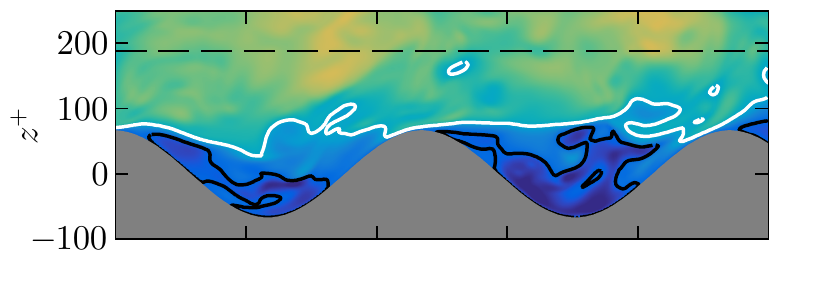}
	\put(-7.1,2.1){(\emph{d})}
	\includegraphics[trim=22 9 0 0,clip = true,scale = 0.88]{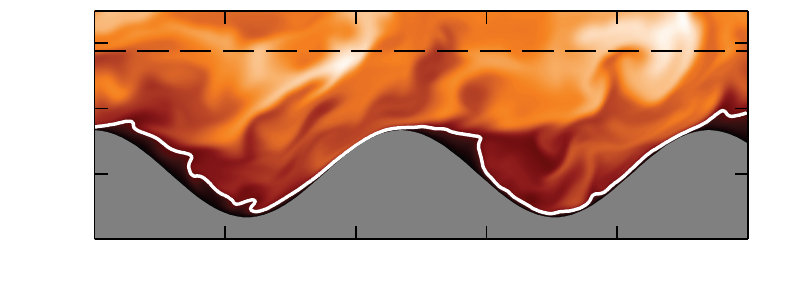}
	\put(-3.82,1.7){\colorbox{white}{$k^+\approx67$, $Re_\tau\approx1200$}}
	\\
	\includegraphics[trim=3 0 5 0,clip = true,scale = 0.88]{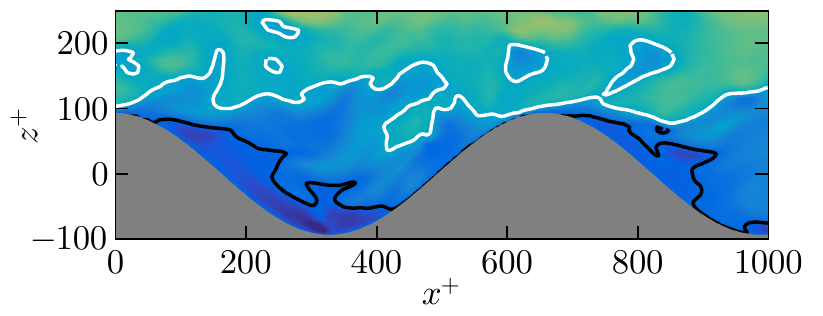}
	\put(-7.1,2.75){(\emph{e})}
	\includegraphics[trim=22 0 0 0,clip = true,scale = 0.88]{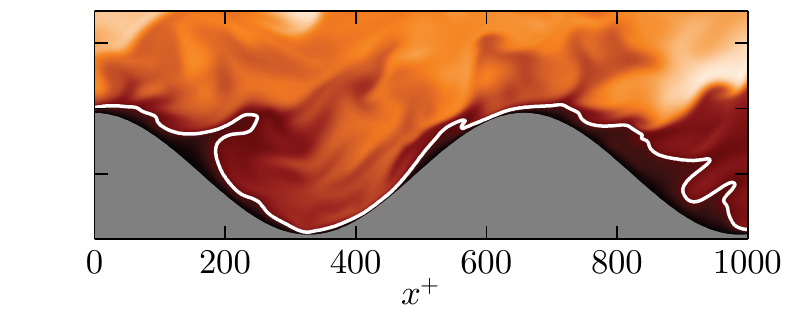}
	\put(-3.82,2.3){\colorbox{white}{$k^+\approx93$, $Re_\tau\approx1680$}}
	\\
	\includegraphics[trim =-25 0 0 0, clip = true]{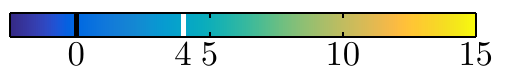}
	\includegraphics[trim =-25 0 0 0, clip = true]{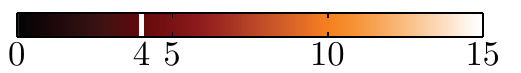}
	\put(-9.1,-0.3){$u^+$}
	\put(-3.4,-0.3){$\theta^+-\theta_w^+$}
	\vspace{+0.1\baselineskip}
\caption{(Colour online) Instantaneous streamwise velocity (left) and temperature (right) in the streamwise--vertical plane, where $\theta_w=0$ is the isothermal wall temperature. Horizontal dashed line shows the minimal channel critical height, $z_c=0.4L_y$. White contour line shows the value $u^+=4$ and $\theta^+-\theta_w^+=4$, to highlight the viscous and thermal diffusive sublayers. Black contour line shows zero velocity (recirculation). Flow is from left to right.
}
	\label{fig:instTemp}
\end{figure}

Figure \ref{fig:instTemp} shows instantaneous snapshots of the streamwise velocity and fluid temperature in the streamwise--vertical plane for increasing roughness sizes. A white contour line for $u^+=4$ and $\theta^+-\theta_w^+=4$ has been selected to provide an indication of the viscous and thermal diffusive sublayers.
The smooth-wall flow (figure \ref{fig:instTemp}\emph{a}) produces thin sublayers close to the wall, for both the velocity and temperature fields.
The roughness in the transitionally rough regime (figure \ref{fig:instTemp}\emph{b},\emph{c}) produces much thicker sublayers. Here, the fluid temperature in most of the region below the roughness crests is near its isothermal wall value, suggesting a reduction in the local temperature gradients at the wall and hence a reduced heat transfer rate in these regions. However, the large temperature gradients  localised to the roughness crests (see also figure \ref{fig:tauPhiContour}\emph{b}) lead to an increase in the overall heat transfer (Stanton number) for these transitional rough-wall cases relative to the smooth wall.
The black contour for the streamwise velocity fields corresponds to zero velocity, giving an indication of the region of recirculation residing behind the roughness elements.
As the roughness height increases further and becomes nominally fully rough (figure \ref{fig:instTemp}\emph{d},\emph{e}) we see that much of the fluid below the roughness crests remains near zero, with the selected contour line of $u^+=4$ residing mostly above the roughness crests, indicating an increasing dominance of pressure drag. 
 In contrast, the thermal diffusive sublayer is seen to be descending into the roughness canopy for these nominally fully rough cases. It appears as a thin sublayer that follows the roughness geometry and resembles that of the smooth wall if the wall was contorted.
 The fluid temperature below the roughness crests is therefore much larger than what might have been expected from inspection of the velocity fields.

These figures, especially the contrast in viscous and thermal diffusive sublayers in figure \ref{fig:instTemp}(\emph{d}) and (\emph{e}), emphasise the dissimilarity of heat and momentum transfer and therefore the breakdown of Reynolds analogy when roughness is present. 
Moreover, the visual similarity of the fully rough thermal diffusive sublayer to that of a contorted smooth wall offers some explanation for why the rough-wall Stanton number (figure \ref{fig:CfCh}\emph{b}) has a similar trend to that of the smooth wall. 
While the present roughness for $\lambda/k\approx 7.1$ has a 17.8\% increase in wetted surface area compared to the smooth wall, this is not enough to explain the 230\% increase in Stanton number that the rough wall produced.
The substantial changes to the overlying flow dynamics produced by the roughness are therefore likely to account for much of this increase. 
The present qualitative description related to the viscous and thermal diffusive sublayers is based on the single roughness geometry studied in this work. It will be interesting to see whether this disparity in sublayers is also observed when the
roughness is varied, for example with densely packed (short wavelength) roughness or with sharp-edged cuboid roughness. While not studied here, these geometries can be readily investigated using the minimal-span channel technique.

%
%
\section{Conclusions}
\label{sect:conc}
Rough-wall turbulent heat-transfer studies have a large parameter space, where the roughness length scales (height,  wavelength and skewness, to name a few) can be varied independently of the flow properties (Reynolds number and Prandtl number). Moreover, the expense of conventional numerical simulations has made exploring this parameter space challenging.
We have demonstrated that the minimal channel can be used to  study the near-wall region of forced convection turbulent flows over three-dimensional sinusoidal roughness. This enables high fidelity data of the near-wall turbulent flow to be simulated with the same level of accuracy as conventional DNS, promising an efficient means to simulate multiple roughness geometries and flow conditions. 

In particular, we have shown that the minimal channel technique can place the fully rough regime of roughness within reach, where the roughness function, $\Delta U^+$, tends towards an asymptote of $\Delta U_{FR}^+=(1/\kappa_m)\log(k^+)+C$. The temperature difference $\Delta \Theta^+$ (\ref{eqn:logT}), meanwhile, was observed to be tending towards a constant value, which for the present sinusoidal roughness was found to be $\Delta \Theta_{FR}^+\approx 4.4$.
With this constant value obtained from the minimal channel, along with the equivalent sand-grain roughness $k_s$, the Stanton number can then be estimated for any roughness Reynolds numbers through  (\ref{eqn:Stanton_FR}). This assumes a logarithmic function for the rough-wall temperature and velocity profiles across the entire channel and resulted in good agreement with the data from our minimal channel simulations. The Stanton number was maximum in the transitionally rough regime, before starting to reduce monotonically in the fully rough regime as $\Delta \Theta^+$ attains a constant value.
The ratio between the momentum and heat transfer roughness lengths, $z_{0m}/z_{0h}$, depends on the Reynolds number (figure \ref{fig:CfCh}\emph{d}). As with the smooth wall, the inner-normalised heat transfer roughness length, $z_{0h}^+$, becomes constant in the fully rough regime, so that the ratio $z_{0m}/z_{0h}$ is linearly proportional to $z_{0m}^+$. The constant of proportionality is related to the roughness-dependent constant $\Delta \Theta_{FR}^+$ through (\ref{eqn:z0h}).

Analysis of instantaneous temperature fields revealed that in the fully rough regime there is a thin thermal diffusive sublayer that follows the roughness geometry, resembling a contorted smooth wall. This is in contrast to the velocity fields, which showed much of the fluid below the roughness crests being close to zero. The differences between these two fields are due to the pressure drag acting on the rough wall, which directly influences the momentum transfer but not the heat transfer. This causes the Reynolds analogy, or similarity between momentum and heat transfer, to break down so that the factor $2C_h/C_f$ is no longer constant (figure \ref{fig:CfCh}\emph{c}). However, when looking beyond the bulk coefficients, a similar distribution pattern between the viscous stress and heat flux at the wall was observed. The reduction of the viscous stress, and hence heat flux, in the fully rough regime explains why the Stanton number reduces and $\Delta \Theta^+$ attains a constant value.

\section*{Acknowledgements}
The authors would like to gratefully acknowledge the financial support of the Australian Research Council through a Discovery Project (DP170102595). This research was supported by computational resources provided from Melbourne Bioinformatics at the University of Melbourne, Australia, and also from the Pawsey Supercomputing Centre with funding from the Australian Government and the Government of Western Australia.


\bibliographystyle{jfm}
\bibliography{bibliography}

\end{document}